\documentclass[twocolumn,trackchanges]{aastex701}
\usepackage{xspace}
\usepackage{amsmath}

\newcommand{\ixpe}{\textit{IXPE}\xspace}

\newcommand{\fermi}{\textit{Fermi}\xspace}
\newcommand{\swift}{\textit{Swift}\xspace}

\newcommand{\xrism}{\textit{XRISM}\xspace}

\newcommand{\agile}{\textit{AGILE}\xspace}

\newcommand{\ixpeobssim}{\textsc{ixpeobssim}\xspace}
\usepackage{comment}

\usepackage{graphicx}
\usepackage{blindtext}

\shorttitle{Stability of Accretion Geometry in \mbox{Cyg X-3}}
\shortauthors{Miku\v{s}incov\'{a} et al.}
\begin{document}

\title{Super-Eddington Accretion Geometry: \\ 
a Remarkable Stability of the Hidden Ultraluminous X-Ray Source Cygnus X-3}

\author[0000-0001-7374-843X]{Romana Miku\v{s}incov\'{a}}
\affiliation{INAF Istituto di Astrofisica e Planetologia Spaziali, Via del Fosso del Cavaliere 100, 00133 Roma, Italy}
\email[show]{romana.mikusincova@inaf.it}

\author[0000-0002-5767-7253]{Alexandra Veledina}
\affiliation{Department of Physics and Astronomy, 20014 University of Turku, Finland}
\affiliation{Nordita, KTH Royal Institute of Technology and Stockholm University, Hannes Alfv\'ens v\"ag 12, SE-10691 Stockholm, Sweden}
\email{alexandra.veledina@gmail.com}

\author[0000-0003-3331-3794]{Fabio Muleri}
\affiliation{INAF Istituto di Astrofisica e Planetologia Spaziali, Via del Fosso del Cavaliere 100, 00133 Roma, Italy}	
\email{fabio.muleri@inaf.it}

\author[0009-0009-3289-3767]{Raul Ciancarella}
\affiliation{Independent Researcher}
\email{raul.cianca@gmail.com} 

\author[0000-0002-0333-2452]{Andrzej Zdziarski}
\affiliation{Nicolaus Copernicus Astronomical Center, Polish Academy of Sciences, Bartycka 18, PL$-$00$-$716 Warszawa, Poland}	
\email{andrzej.zdziarski@gmail.com}

\author[0000-0003-3189-9998]{David A.\ Green}
\affiliation{Astrophysics Group, Cavendish Laboratory, J. J. Thomson Avenue, Cambridge CB3 0US, UK}
\email{dag@mrao.cam.ac.uk}

\author[0000-0002-8384-3374]{Michael McCollough}
\affiliation{Harvard-Smithsonian Center for Astrophysics, 60 Garden Street, Cambridge, MA 02138, USA}
\email{mmccollough@cfa.harvard.edu}

\author[0000-0002-1084-6507]{Henric Krawczynski}
\affiliation{Physics Department and McDonnell Center for the Space Sciences, Washington University in St. Louis, St. Louis, MO 63130, USA}
\email{krawcz@wustl.edu}

\author[0000-0002-5872-6061]{James~F.~Steiner}
\affiliation{Center for Astrophysics, Harvard \& Smithsonian, Cambridge, MA 02138, USA}
\email{james.steiner@cfa.harvard.edu}

\author[0000-0003-0079-1239]{Michal Dov\v{c}iak}
\affiliation{Astronomical Institute of the Czech Academy of Sciences, Bo\v{c}n\'{i} II 1401/1, 14100 Praha 4, Czech Republic}
\email{michal.dovciak@asu.cas.cz} 

% ...

%\end{comment}

%\collaboration{all}{The Terra Mater collaboration}

\begin{abstract}
We report on the average and orbital phase-resolved polarization of \mbox{Cyg X-3} in the hard state during the 2023 Imaging X-ray Polarimetry Explorer (IXPE) observational campaign. 
We find the polarization degree of $ 21.2 \pm 0.4 \% $ and polarization angle of $ 92\fdg2\pm0\fdg5 $, well compatible with the first hard-state IXPE observation in 2022.
As the observed polarization depends on both the accretion geometry and the X-ray emission mechanism, which we attribute to reflection from the optically thick envelope surrounding the central source, our result indicates that both are very stable on year-long timescale.
We discuss time- and energy-dependent polarization properties and their implications for the geometry and stability of the accretion funnel.
\end{abstract}



\section{Introduction} 

Cygnus X-3 (hereafter \mbox{Cyg X-3}) is a persistent radio, X-ray and $\gamma$-ray source, and a high-mass X-ray binary system composed of a compact object, likely being a neutron star or a low-mass black hole (BH) \citep{Zdziarski2013}, and a Wolf-Rayet (WR) star \citep{Kerkwijk1996}. The findings of \cite{Zdziarski2013} of a low-mass BH are different from those of \cite{Antokhin2022}, who favored $ \sim 7~$M$ _{\odot} $. However, a new research based on \xrism results, Miura+25 (PASJ, in press, 2505.09890) find the results very similar to the earlier findings.
The orbital period of \mbox{Cyg X-3} is $P \approx 4.8$~h \citep{Parsignault1972}, which is among the shortest known for X-ray binaries.
Interestingly, since its discovery \citep{Giacconi1967}, the source has been known to exhibit peculiar multiwavelength properties.
In fact, \mbox{Cyg X-3} has attracted particular interest since the first observation of its radio flares \citep{Gregory1972}, when it becomes the brightest X-ray binary in radio with flux density reaching 20 Jy \citep[e.g.,][]{McCollough1999, Corbel2012, Spencer2022}.

The source is also atypically bright in the $\gamma$-rays \citep{Atwood2009, Tavani2009}, suggesting that a peculiar accretion geometry might be favoring the conditions for launching truly powerful outflows. \mbox{Cyg X-3} exhibits transient $\gamma$-ray emission above 100 MeV, firmly detected by \agile~ and \fermi/LAT \citep{Tavani2009, Abdo2009}. These $\gamma$-ray episodes occur when the system is in the soft X-ray spectral state and just before major radio flares, with the peak isotropic luminosity reaching $ \sim 10^{36}$~erg~s$^{-1}$, assuming a distance of 7--10 kpc. More recent findings by \cite{Reid2023} constrain the distance to 9--10 kpc. The $\gamma$-ray spectrum is well described by a power law ${\rm d}N/{\rm d}E\propto E^{-\Gamma}$ with a photon index $\Gamma\sim2$  \citep{Piano2012}, and the emission is modulated on the 4.8-h orbital period, suggesting an origin in the relativistic jet at distances of $\sim 10^{10}$--$\sim 3 \times 10^{12}$~cm from the compact object \citep{Dubus2010, Cerutti2011}. The emission mechanism is likely leptonic, produced via inverse Compton scattering of UV photons from the WR companion star by relativistic electrons in the jet, although hadronic scenarios remain possible \citep{Romero2003, Zdziarski2012a}. The source has indeed been associated to neutrino emission \citep{Koljonen2023}.

X-ray polarimetric studies, enabled by the Imaging X-ray Polarimetry Explorer \citep[IXPE;][]{Weisskopf2022, Soffitta2021}, offer a novel diagnostic that can resolve longstanding uncertainties about the geometry of the emission region. 
IXPE can measure the linear polarization in the 2\--8~keV energy range, providing space-resolved (with $\sim$30~arcsec half-power diameter) and energy-resolved (resolution better than 20\% at 6 keV) observations.
Several X-ray binary systems \citep{Dovciak2024} were found to have high, between about 3\% and 12\%, polarization degree (PD) in the hard spectral state: Cyg~X-1 \citep{Krawczynski2022, Kravtsov2025}, Swift~J1727.8$-$1613 \citep{Veledina2023, Ingram2024, Podgorny2024} and IGR~J17091$-$3624 \citep{Ewing2025}, with the polarization angle (PA) coinciding with the position angle of the jet (which is also aligned with the optical polarization).
These results have been interpreted in terms of the hot, X-ray emitting medium being elongated in the disk plane.
The hard-state (HS) geometry remains stable despite X-ray flux variations spanning up to two orders of magnitude \citep{Podgorny2024, Kravtsov2025}. The results for Cyg~X-1 and IGR~J17091$–$3624 may require the Comptonizing medium to move at half the speed of light to explain the high PD values \citep{Beloborodov1998, Poutanen2023, Krawczynski2025}.

\begin{deluxetable*}{ccccccc}
\tablecaption{Observations of \mbox{Cyg X-3} performed by \ixpe and its measured polarimetric properties in their respective spectral states. Uncertainties at 1$ \sigma $ CL.}
\tablenum{1}
\tablehead{\colhead{Obs ID} & \colhead{Start Date} & \colhead{Exposure} & \colhead{Spectral state} & \colhead{PD} & \colhead{PA} & \colhead{ID} \\ 
\colhead{} & \colhead{} & \colhead{[ks]} & \colhead{} & \colhead{[\%]} & \colhead{[deg]} & \colhead{} } 

%% All data must appear between the \startdata and \enddata commands
\startdata
02001899  &   2022-10-14      & 538  &  HS   & $20.6 \pm 0.3$  & $90.1 \pm 0.4$ & HS 2022  \\ 
02250301  &   2022-12-25      & 199  &  IMS  & $10.4 \pm 0.3$  & $92.6 \pm 0.7$ & IMS 2022  \\ 
02009101  &   2023-11-17      & 291  &  HS   & $21.4 \pm 0.4$  & $92.2 \pm 0.5$ & HS 2023  \\ 
03250301  &   2024-06-02      & 50   &  SS   & $11.9 \pm 0.4$  & $94.0 \pm 1.0$ & SS 2024  \\ 
\enddata
\label{tab:ObsID}
\tablecomments{Observation HS 2023 is the main focus of this paper. HS 2022 and IMS 2022 are discussed in \cite{Veledina2024}, SS 2024 in \cite{Veledina2024b} and Rodriguez-Cavero et al. (in prep.). We define the reference ID for each observation in the last column and adopt this notation throughout the paper.}

\end{deluxetable*}

The spectral states of \mbox{Cyg X-3} are similar to other X-ray binary systems harboring BHs \citep{Szostek2008}. 
The source switches spectral states in a known, repeating manner, with the X-ray spectral states being tightly related to the radio properties.
It is most often found in the hard X-ray, quiescent radio state, in which the X-ray spectrum is dominated by a power-law-like continuum with a prominent reflection complex \citep[e.g.,][]{Hjalmarsdotter2008}, and the flux density in radio bands are $\sim$100~mJy \citep[e.g.,][]{McCollough1999, Trushkin2017}.
Incursions into the X-ray intermediate state (IMS) are accompanied by an increase of radio flux densities up to $\sim$0.3~Jy and softening of the X-ray spectrum.
The intermediate state is sometimes followed by a transition to the ultrasoft X-ray (SS), quenched radio state, with a blackbody-like spectrum and low levels of radio flux density $\lesssim$10~mJy.
After that, the source shows major radio ejections and closes the activity loop by returning to the hard state. Each spectral state typically lasts from several weeks to a few months, as observed in long-term monitoring with the \swift/BAT observatory.

The physical mechanism of the repeating spectral state changes observed over long time scales in X-ray and radio properties of \mbox{Cyg X-3} is not known.
Potential scenarios include changes of the local absorption \citep{Hjalmarsdotter2009}, the transformation of the funnel-like accretion geometry into disk-dominated accretion, akin to typical low-mass BH X-ray binaries in the soft spectral state, or the repeating filling of an inner funnel with matter and its subsequent clearing \citep{Veledina2024b}.
Recent \ixpe polarimetric observations have indeed revealed evidence for an atypical geometry in \mbox{Cyg X-3} \citep{Veledina2024, Veledina2024b}.
The distinct hard-state spectra of the source, closely resembling reflected emission, and the exceptionally high polarization degree (PD) $\sim$20\% orthogonal to the direction of radio ejections are reminiscent of those detected in obscured Seyfert 2 galaxies \citep[e.g.,][]{Ursini2023}. These properties favor a geometry where the source of the incident X-ray emission is covered by a thick conical structure located high above the orbital plane. At an inclination of 30\degr \citep{Veledina2024, Veledina2024b}, this structure blocks direct radiation from the line of sight, and the observer sees only reflected emission from its inner walls.
The inferred luminosity, both intrinsic and apparent for the observer looking down the funnel, appeared to exceed $3\times10^{39}$~erg~s$^{-1}$, rendering \mbox{Cyg X-3} in the class of ultraluminous X-ray sources (ULXs) and then the first source of this kind studied through X-ray polarimetric analysis.

A clue to this puzzle is an important constraint of accretion geometry: if a substantial fraction of the funnel material is blown away together with the major radio ejection, one may expect the gradual accumulation of matter during episodes of quiescent radio emission, until the subsequent ejection.
Parameters of the funnel, such as the size, optical thickness and opening angle, determine the beaming factor and hence the brightness of this off-axis ULX.
Variations in the opening angle would manifest themselves in changes of the apparent luminosity, while the size of the funnel proportionally affects the amount of observed radiation. Previous findings by \cite{Veledina2024} constrained the opening angle in the hard state $ \lesssim 15\degr $.

In this paper, we explore the long-term evolution of the accretion geometry of \mbox{Cyg X-3} in the hard X-ray spectral state.
\ixpe observed \mbox{Cyg X-3} in November 2023, about one year after the first hard-state observation of the source.
In this paper, we report on the observed average and orbital phase-resolved polarimetric properties of the source.
We describe the data reduction in Section~\ref{sec:methods}. 
The results of data analysis and their implications for the properties of the system are presented in Section~\ref{sec:results}.
We summarize our findings in Section~\ref{sec:summary}.
In line with current consensus, in this work, we assume that \mbox{Cyg X-3} harbors a BH.

\begin{figure*}%[ht!]
\plotone{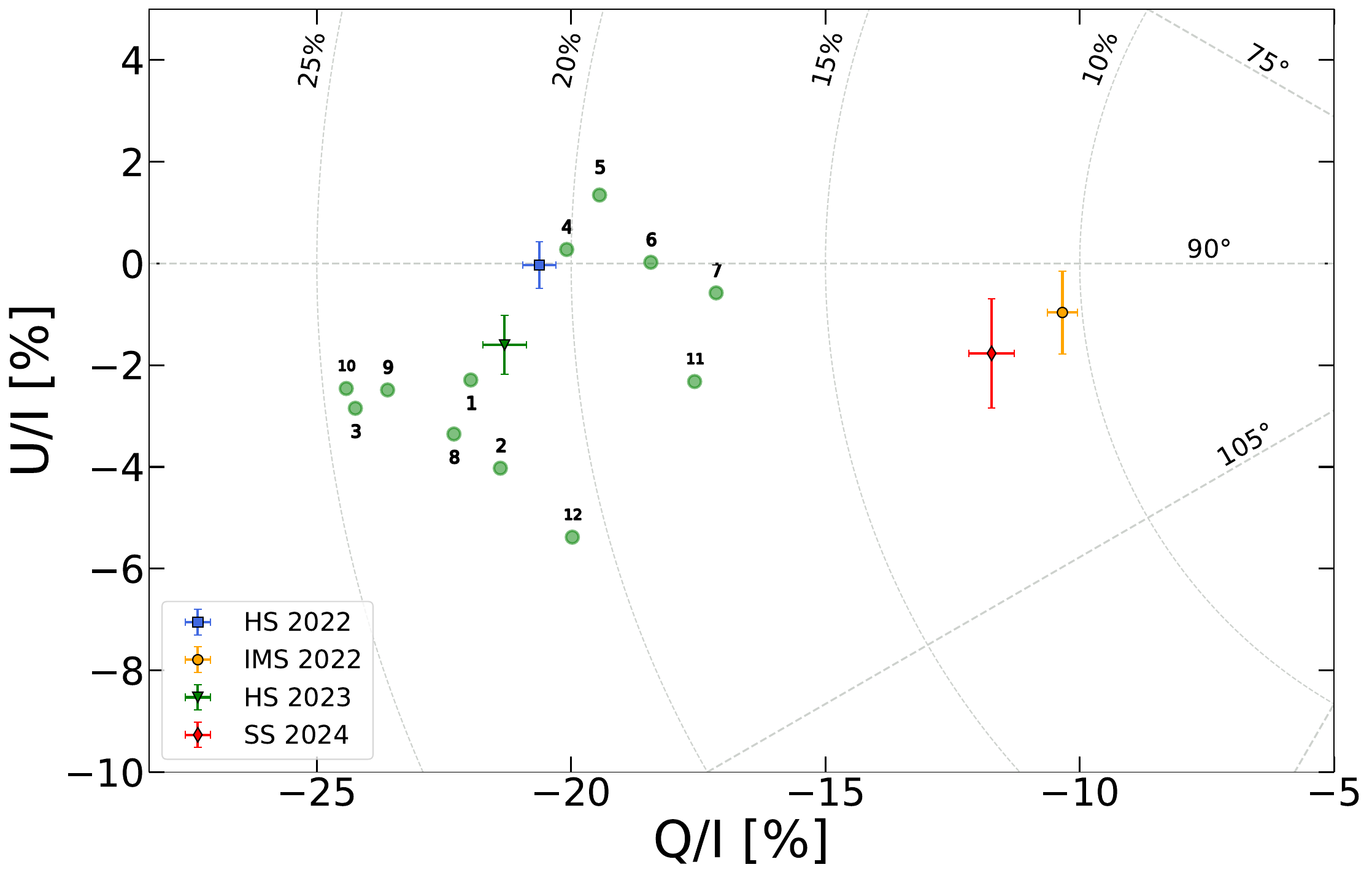}
\caption{Normalized Stokes parameters $U/I$ versus $Q/I$ for the \ixpe observations of \mbox{Cyg X-3}. HS 2022 in blue, IMS 2022 in yellow, HS 2023 in green and SS 2024 in red color. The error bars are plotted at $1\, \sigma$ CL. 
In the case of HS 2023, we also plot the change of the polarization properties with the orbital phase in fainter green, numbered 1--12, with bin 1 centered on phase 0. The light green points are shown without error bars, which are on average $\Delta (Q/I)   \approx  \Delta (U/I)\approx 2\%$ ($1\, \sigma$). The gray grid represents the PD (in \%) and PA (in $ \degr $).
\label{fig:pcube}}
\end{figure*}

\section{Observation and Methods} \label{sec:methods}

\mbox{Cyg X-3} was observed by \ixpe during 4 epochs in total so far.
All of the observations are reported in Table~\ref{tab:ObsID}. Over these epochs, the source was caught in three spectral states showing different polarization properties. The hard spectral state (HS) shows prominent PD of $\sim$20\%, higher than the ultrasoft (SS) and intermediate (IMS) spectral states. The latter two differ by only $\Delta$PD$\sim$1.5\%. On the other hand, PA stays constant within 3$\sigma$ between all of the observations.

%Fig 2.
\begin{figure}

\includegraphics[width=.495\textwidth]{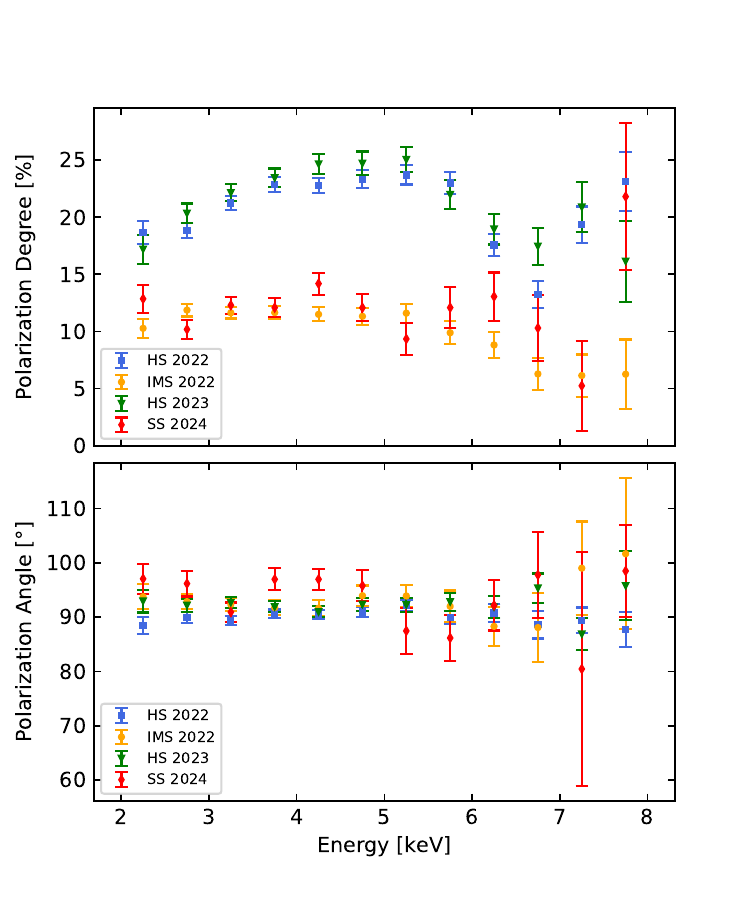}
\caption{Dependence of PD (top) and PA (bottom) on energy in the 2--8 keV band. HS 2022 in blue squares, IMS 2022 in yellow circles, HS 2023 in green triangles and SS 2024 in red diamonds.
\label{fig:energy}}
\end{figure}

Here we focus on the third observation of \mbox{Cyg X-3} (Observation ID 02009101, hereafter referred to as HS 2023), which took place between 2023 November 17 (start time 20:16:16 UTC) and 2023 November 23 (end time 23:08:15 UTC) for approximately 291~ks, and compare it with the other three \ixpe observations of the source.
We downloaded the Level 2 data from the public HEASARC data archive and proceeded with the data reduction. 
The \texttt{barycorr} tool, from the \textsc{ftools} package (\textsc{heasoft} software version 6.35.1), was applied to the data in order to adjust the timing of the events' arrival relative to the solar system's barycenter. We used the \ixpeobssim software (v. 31.0.1; \citealt{Baldini2022}) and its \texttt{xpselect} feature to select the source region. The extraction regions were circular with a radius of 90\arcsec. Since the source is very bright ($ > $ 2 count s$^{-1}$), no background rejection or subtraction was  necessary \citep{DiMarco2023}. The \texttt{xpbin} tool of \ixpeobssim was used to create the polarization cubes (\texttt{pcube}) as well as to generate the $I$, $Q$, and $U$ Stokes spectra. In the end, the \texttt{ftgrouppha} (\textsc{ftools}) was used to rebin the Stokes spectra to contain 8 counts per bin for Stokes $I$, $Q$, and $U$. 
Version 20230101\_v013 of the response matrices was used to create these products, applying unweighted (\texttt{pcube}) and weighted (Stokes $I$, $Q$, $U$) analysis \citep{DiMarco2022}.

It is well-established that the source undergoes pronounced flux modulation with the orbital phase of the binary system \citep{Antokhin2022}. 
Also, the source showed orbital variations of the X-ray polarization during the first two observations carried out by \ixpe 
\citep{Veledina2024}.

Data from HS 2023 are therefore phase-folded based on the quadratic ephemeris model presented in Table 2 of \citet{Antokhin2019}, where phase 0 corresponds to the superior conjunction of the system.

We then calculated the phase of each detected event and added this information to the data file. 
We split the observation into 12 phase bins, with phase 0 defined as the center of the first bin.

% Figure 3
\begin{figure}
\centering
\includegraphics[width=0.495\textwidth]{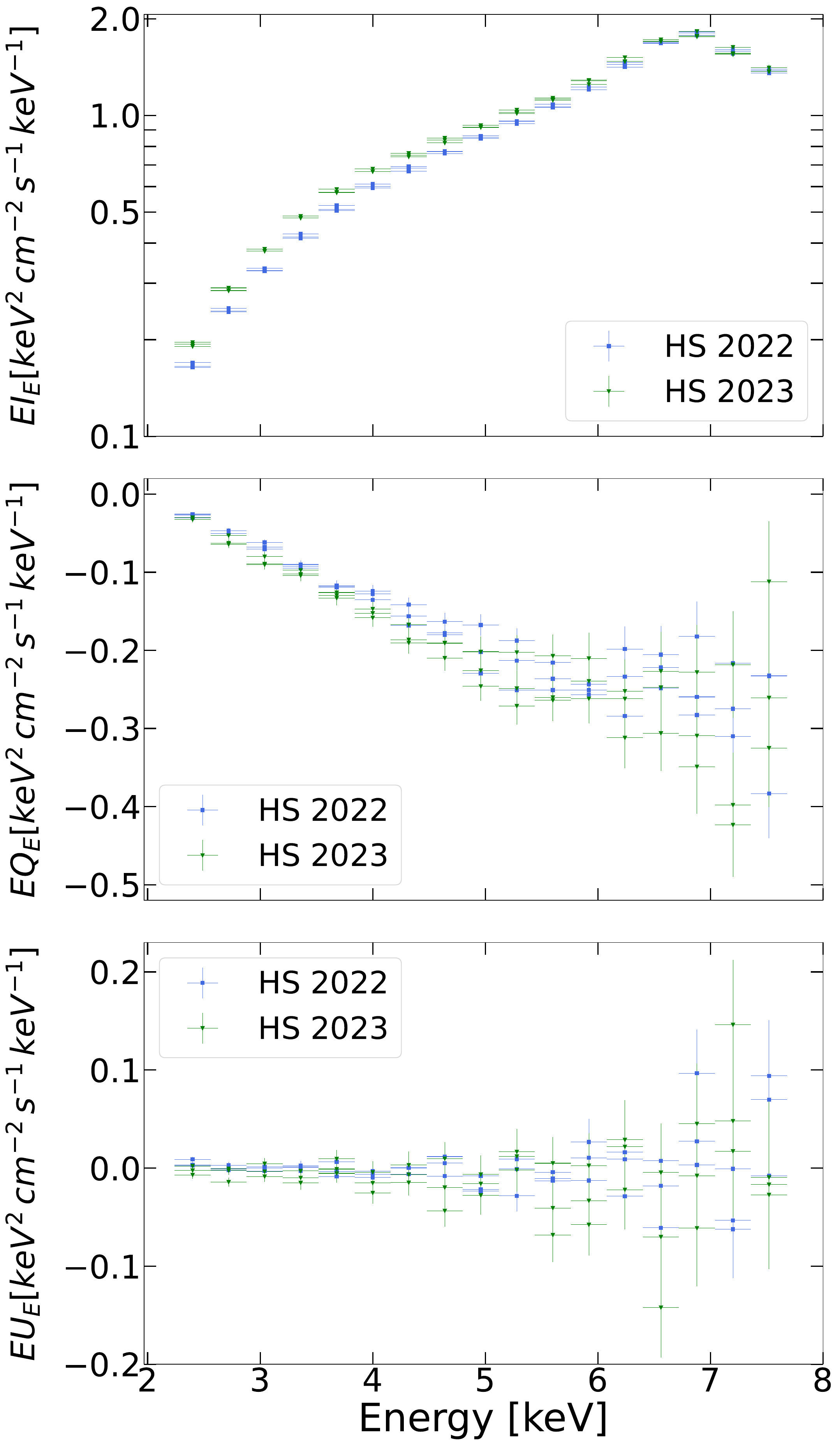}
\caption{Top to bottom: unfolded $EF_E$ (energy flux) spectra of Stokes $I$, $Q$, and $U$, respectively, comparing the spectral shape between HS 2022 (blue) and HS 2023 (green). Spectra are unfolded against a power law with photon index 1.7, for plotting purposes only.
\label{fig:stokes}}
\end{figure}

\section{Results} \label{sec:results}

\subsection{Energy-dependent polarimetry}

Using the \texttt{pcube} algorithm of \ixpeobssim to determine the presence of model-independent polarization, we obtain the values of energy-averaged PD and PA, which are depicted in Figure \ref{fig:pcube}. During HS 2023, the source shows high PD=$21.4\pm0.4$\% at  PA=$ 92\fdg2\pm0\fdg5 $ (hereafter errors are quoted at the 68.3\% confidence level). 
Over the \ixpe\ observations of \mbox{Cyg~X-3} conducted so far, the PA has shown almost no variation, whereas the PD has exhibited substantial changes (see Table~\ref{tab:ObsID}). 
There is a difference of $\Delta$PD$\sim$10\% between the PD in the hard state (HS 2022 and HS 2023) versus intermediate (IMS 2022) and ultrasoft (SS 2024) states. Interestingly, the intermediate and ultrasoft states differ by only $\Delta$PD$=$1.5$\pm$0.5\%. 
The PD values in HS 2022 and HS 2023 differ by $\Delta$PD$=0.8\pm 0.5\%$ ($1\, \sigma$).

\begin{figure} %[h!]
\includegraphics[width=1.\linewidth]{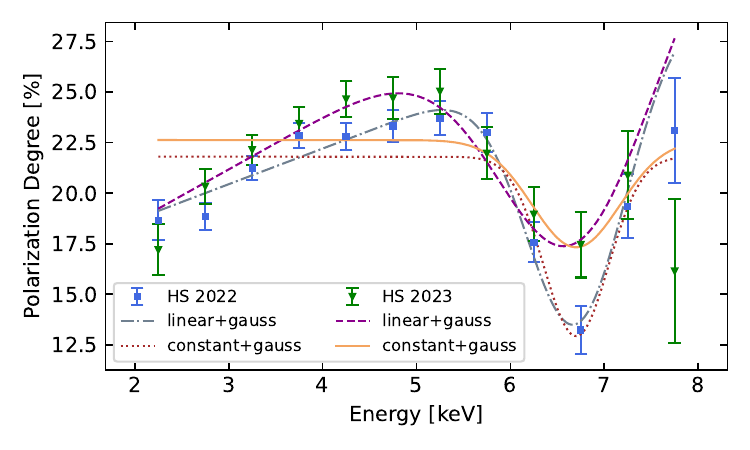}
\caption{IXPE PDs (data points, HS 2022 in blue and HS 2023 in green) with two PD models fitted to the data of the two epochs: a linear change of the PD with energy plus a non-polarized gaussian is shown with the dash-dotted gray line (HS 2022) and the dashed purple line (HS 2023). A constant PD model with a non-polarized Gaussian is shown with the dotted brown line (HS 2022) and the solid orange line (HS 2023).}
\label{fig:PDfit}
\end{figure}

In Figure \ref{fig:energy} we show the energy dependence of the polarization properties for the four observations. Over the entire 2--8 keV waveband, both PD and PA do not exhibit large variations, although some minor changes are observed, especially in the case of PD. In particular, there is a slight increase in PD in the energy range 2--5 keV followed by a moderate decrease as we reach towards 8 keV (HS 2022 and HS 2023) and is due to the presence of prominent iron lines. 
In a recent paper using high resolution \xrism spectroscopy, \cite{Audard2024} has confirmed this to be a group of fluorescence lines caused by different ionization states of the element, which are expected to be unpolarized. 
The initial PD growth with energy does not seem to be present in the intermediate and ultrasoft states. 
The PA shows a more constant trend across the \ixpe energy band in all four of the observations.

The energy dependence of the Stokes parameters $I$, $Q$, and $U$ in the two hard-state observations, HS 2022 and HS 2023, is shown in Figure~\ref{fig:stokes}. The main difference between the two observations is seen from the spectral shape of the Stokes $I$ datasets, particularly at lower energies, where we notice a softer spectra in the case of HS 2023.

\subsection{Polarization in the hard spectral state}

Although the polarization of the source in HS 2023 is comparable to that of HS 2022, they are not identical.
The difference can be noticed in the spectral dependence of the PD (Figure~\ref{fig:energy}) found by fitting the PD data obtained from the \texttt{pcube} using models with a linear energy-dependent PD and a non-polarized gaussian versus constant polarization with a non-polarized gaussian  (Figure~ \ref{fig:PDfit}). The peak of the gaussian was fixed at 6.7~keV, as that is the energy of Fe~\textsc{xxv} -- the iron type with the highest equivalent width found in the \mbox{Cyg X-3} spectra \citep[Extended Data Figure 6 of][]{Veledina2024}. The fit with linear energy-dependent PD is significantly favored, with a $\chi^{2}$/d.o.f. = 10.2/8 (HS 2022) and a $\chi^{2}$/d.o.f. = 14.8/8 (HS 2023), compared to $\chi^{2}$/d.o.f. = 47.0/9  (HS 2022) and $\chi^{2}$/d.o.f. = 44.3/9  (HS 2023) for the constant PD model.
These results correspond to improvements of $\approx 4\, \sigma$ ($p$-value of $6 \times 10^{-4} $) and $\approx 3.5\, \sigma $ ($p=4 \times 10^{-3}$), respectively. 
This could indicate a suppression of polarization at low energies in HS 2023, which may come from the joint contribution of abundant emission lines \citep[seen in {\it Chandra},][]{Kallman2019}, the unpolarized contribution of the reprocessed emission from the funnel walls, or by the scattered light from the central source within the funnel (polarized at $\sim$10\%; \citealt{Veledina2024b}).

\begin{figure}
\centering 
\includegraphics[width=.495\textwidth]{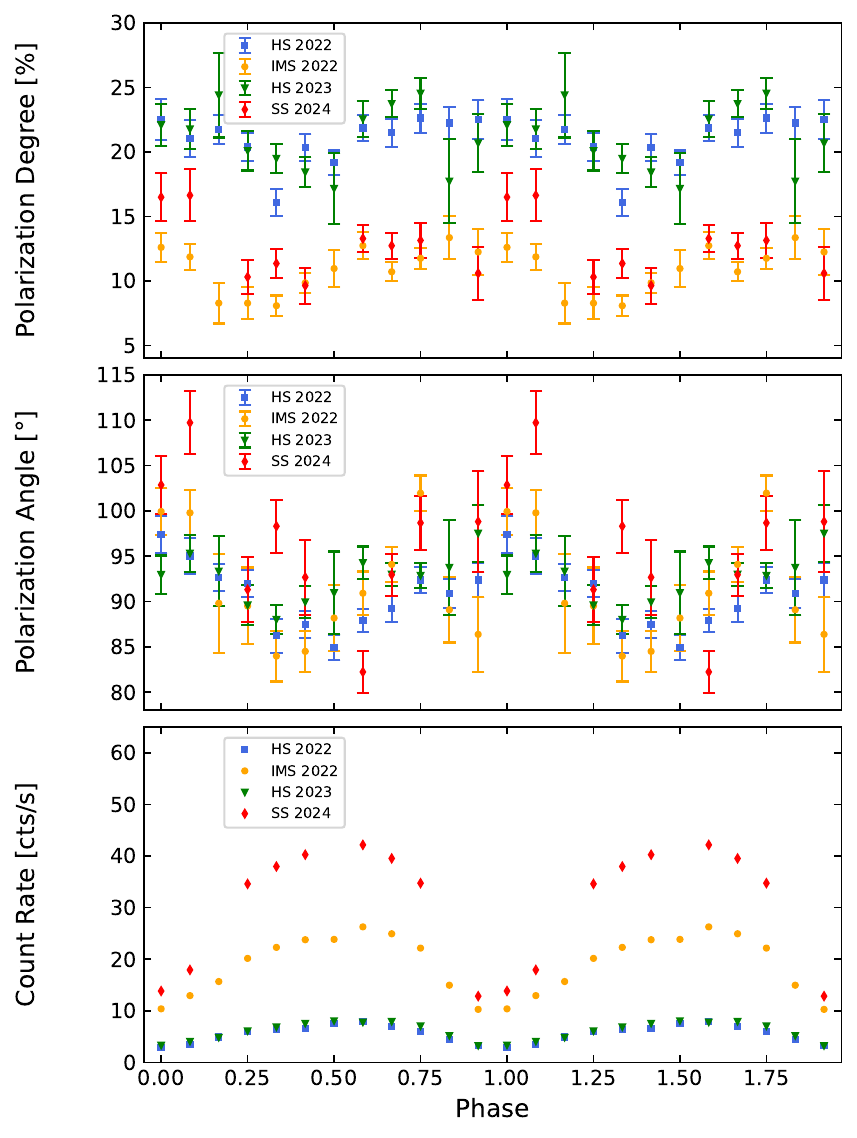}
\caption{Dependence of the PD (top), PA (middle), and count rate (bottom) on the orbital phase. Each data point represents the mean value for a given orbital phase bin. HS 2022 in blue, IMS 2022 in yellow, HS 2023 in green and SS 2024 in red.
\label{fig:phase}}
\end{figure}

\subsection{Orbital phase-resolved polarimetry}

A more intriguing view is offered when we examine the variation of polarization characteristics with the orbital phase of the source. Figure \ref{fig:phase} demonstrate connection between variation of polarimetric properties and count rate. The figure shows a strong orbital modulation of the PD and PA, similar to the previous result of \cite{Veledina2024}. The flux of the source varies systematically across its 4.8 h orbital period. We also find notable changes in the polarimetric properties that are present in all four of the observations of \mbox{Cyg X-3} performed by \ixpe, having a cosine-like behavior. Both results are consistent with the findings of \cite{Antokhin2022}. They showed that the X-ray light curves displayed a deep primary minimum at the superior conjunction of the compact object, followed by an asymmetric recovery and a secondary dip near phase $\approx 0.4$.

The PD of HS 2023 varied between ${\sim}17$\% and ${\sim}25$\%, while PA underwent a modulation between ${\sim}87\degr$ and ${\sim}95\degr$, coinciding with the variations observed during the HS 2022 within the margin of error. These findings are further reinforced by a goodness-of-fit comparison between the constant versus cosine fits. The cosine function only has two free parameters (amplitude and offset), while the angular frequency is defined by the orbital period of the source, and the initial phase is fixed to 0. Seven out of eight datasets (one dataset of PD and one of PA, per each observation) show strong evidence ($\geq 3\, \sigma$) that the cosine model significantly outperforms the constant model. The PD during HS 2023 and the PA during IMS 2022 both show orbital variations at $4.1\, \sigma$ significance. 
The only case showing a moderate improvement in the fit at $2.3\, \sigma$ significance is the PA during SS 2024.
The cosine-like modulation is consistent across all data sets, suggesting a genuine physical effect rather than statistical fluctuation.

\subsection{Radio-X-ray correlation}

\begin{figure}%[ht!]
\centering
\includegraphics[width=0.8\linewidth]{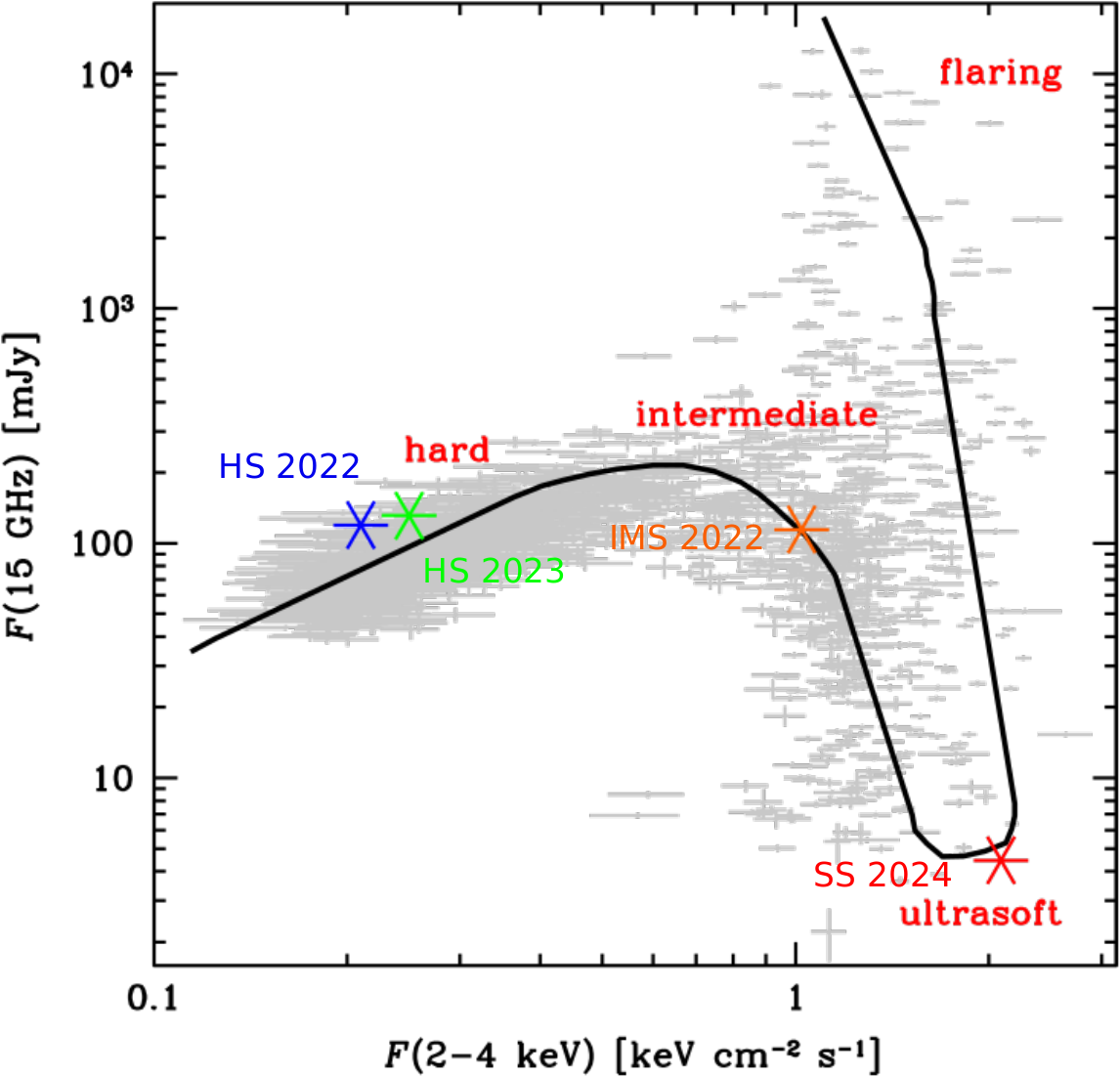}
\caption{X-ray versus radio flux density diagram. Blue, orange, green, and red asterisks indicate the \ixpe observations HS 2022, IMS 2022, HS 2023, and SS 2024, respectively. Grey points represent data from AMI and were presented in \cite{Zdziarski2016}.
\label{fig:CC}}
\end{figure}

In order to better understand the spectral properties of the source, we looked at the X-ray versus radio flux density diagram (Figure~\ref{fig:CC}, using radio data from the Arcminute Microkelvin Imager, presented in \citealt{Zdziarski2016}) during the time of the \ixpe observations. The source is typically found in the hard spectral state, as was the case for two out of the four \ixpe observations (HS 2022 and HS 2023). It is not surprising, although somewhat rare, to see the source in an ultrasoft state. Such a transition is typically accompanied by a rise in the soft X-ray flux and a decrease of radio emission (IMS 2022 and SS 2024). Unlike the case of the polarimetric properties, where we have seen HS 2022 and HS 2023 to be almost identical, the spectral properties of these two observations are more diverse (e.g., $\Delta~count~rate \leq 30 \% $, or the spectral shape becomes notably different below $ \sim $ 5 keV, as shown in the top panel of Figure \ref{fig:stokes}).

\subsection{Time-dependent orbit-average analysis}

\begin{figure*}%[ht!]
\plotone{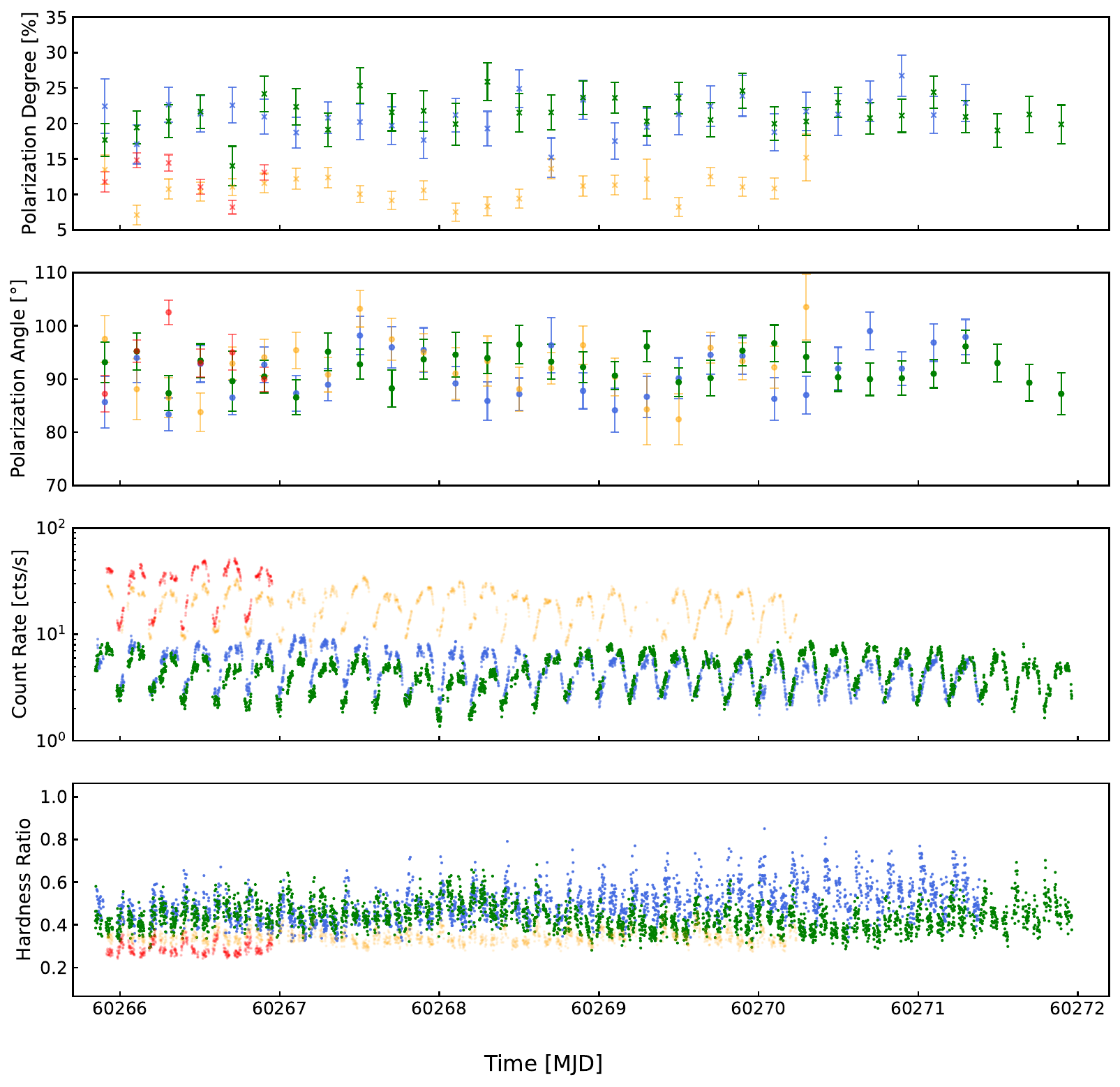}
\caption{Polarization degree, polarization angle, count rate and hardness ratio (top to bottom, respectively) for HS 2022 (blue), IMS 2022 (yellow), HS 2023 (green), and SS 2024 (red) variation with time. HS 2022, IMS 2022 and SS 2024 are shifted for the first superior conjunction of the system during the \ixpe observations to coincide with that of HS 2023. The entire 2--8 keV energy band was considered with data binned in intervals of 500 s.
\label{fig:LC}}
\end{figure*}

We also study the temporal variations in the polarimetric and spectral properties, shown in Figure \ref{fig:LC}. The HS 2022 (blue), IMS 2022 (yellow), HS 2023 (green), and SS 2024 (red) observations were plotted so that their first superior conjunctions are aligned. For all four of these observations, we see orbital phase variations of the polarization properties, with PD of IMS 2022 consistently below the PD of HS 2022 and HS 2023 during the entirety of the observation, while the PA stays stable between all four observations (the first and second panels from the top of Figure~\ref{fig:LC}, respectively). The light curve (third panel from the top) clearly shows orbital variability over the observational period for all four observations. The orbit-average values show flux in the intermediate and ultrasoft states higher than in the two hard states by approximately an order of magnitude.

\section{Summary}\label{sec:summary}

We find that both the time-average and orbital phase-resolved polarimetric properties of HS 2023 are remarkably similar to those of HS 2022, despite the fact that the observations were taken 11 months apart with a flux difference of $\Delta~count~rate \leq 30 \% $ and after the source had made an incursion into the intermediate spectral state (Figures~\ref{fig:pcube} and \ref{fig:phase}).
Based on this finding we can conclude that the geometry of the system is stable for these two \ixpe observations, and, by extension, generally in the hard X-ray spectral state of the system: the primary X-ray source is covered by an optically thick envelope, and the emission escapes through the narrow funnel with the opening angle ${\lesssim}15\degr$, as was found for the HS 2022 \citep{Veledina2024}.

X-ray polarimetry observations of Cyg X-3 reveal that while the polarization properties in the HS 2022 and HS 2023 states are comparable, they show different spectral dependencies, with linear energy-dependent polarization models favored over constant polarization (improvements of $ \sim 4 \sigma $ and $ \sim 3.5 \sigma $, respectively). The data suggest a possible suppression of polarization at low energies observed in HS 2023, which could potentially result from contributions of emission lines, unpolarized reprocessed emission from funnel walls, or scattered light from the central source.

We find the HS 2023 orbit-average time-dependent polarization is constant, similar to HS 2022 (Figure~\ref{fig:LC}).
This contrasts with the higher spread of orbit-average values of PD in the intermediate and ultrasoft states, where we find marginal significance for time-dependent variations.
This finding suggests that the polarization production mechanism remains stable in the HS and the geometry is not too sensitive to parameter variations.
However, we note that only a single observation for each of the intermediate and soft states has been carried out with \ixpe. Additional polarimetric observations in these states are needed to assess whether their geometries are similarly stable over time. Furthermore, observations of the source in the flaring state—the only spectral state not yet observed with \ixpe—would provide valuable insights into the polarization behavior across the complete range of accretion states.
This bolsters the earlier finding by \cite{Veledina2024} that a small, ${\lesssim}15\degr$, funnel opening angle is preferred over a larger ${\sim}30\degr$ funnel opening angle. 
This is because, although both opening angles give the same polarization, in the latter case the PD is highly sensitive to variations in the opening angle on the scale of less than a degree \citep[figure 4b of][]{Veledina2024}.
While the influence of the funnel size and opening angle on the observed emission is now established, the role of other physical parameters of the funnel remains uncertain and requires further observational constraints.

%% Please use the acknowledgment and contribution environments. This will 
%% be anonomyized when the "anonymous" style option is used. 
\begin{acknowledgments}
The Imaging X-ray Polarimetry Explorer (IXPE) is a joint US and Italian mission.  The US contribution is supported by the National Aeronautics and Space Administration (NASA) and led and managed by its Marshall Space Flight Center (MSFC), with industry partner Ball Aerospace (contract NNM15AA18C). The Italian contribution is supported by the Italian Space Agency (Agenzia Spaziale Italiana, ASI) through contract ASI-OHBI-2022-13-I.0, agreements ASI-INAF-2022-19-HH.0 and ASI-INFN-2017.13-H0, and its Space Science Data Center (SSDC) with agreements ASI-INAF-2022-14-HH.0 and ASI-INFN 2021-43-HH.0, and by the Istituto Nazionale di Astrofisica (INAF) and the Istituto Nazionale di Fisica Nucleare (INFN) in Italy.  This research used data products provided by the IXPE Team (MSFC, SSDC, INAF, and INFN) and distributed with additional software tools by the High-Energy Astrophysics Science Archive Research Center (HEASARC), at NASA Goddard Space Flight Center (GSFC). 

A.V. acknowledges support from the Academy of Finland grant 355672. Nordita is supported in part by NordForsk.
AAZ acknowledges support from the Polish National Science Center
grants 2019/35/B/ST9/03944 and 2023/48/Q/ST9/00138.
M.D. and J.S. thank GACR project 21-06825X for the support and institutional support from RVO:67985815. 
The work of G.M. is partially supported by the PRIN 2022 - 2022LWPEXW - “An X-ray view of compact objects in polarized light”, CUP C53D23001180006.
A.I. acknowledges support from the Royal Society. 
P.O.P. acknowledges financial support from the French Space National Agency (CNES) and the National Center of Scientifc Research (CNRS) via its "Action Thématique" PEM. J.Pod. acknowledges institutional support from RVO:67985815. IL was funded by the European Union ERC-2022-STG - BOOTES - 101076343. Views and opinions expressed are however those of the author(s) only and do not necessarily reflect those of the European Union or the European Research Council Executive Agency. Neither the European Union nor the granting authority can be held responsible for them.
\end{acknowledgments}

\bibliography{sample701}{}

\begin{thebibliography}{}
\expandafter\ifx\csname natexlab\endcsname\relax\def\natexlab#1{#1}\fi
\providecommand{\url}[1]{\href{#1}{#1}}
\providecommand{\dodoi}[1]{doi:~\href{http://doi.org/#1}{\nolinkurl{#1}}}
\providecommand{\doeprint}[1]{\href{http://ascl.net/#1}{\nolinkurl{http://ascl.net/#1}}}
\providecommand{\doarXiv}[1]{\href{https://arxiv.org/abs/#1}{\nolinkurl{https://arxiv.org/abs/#1}}}

% type= article
\bibitem[{A.~A. {Abdo} {et~al.}(2009){Abdo}, {Ackermann}, {Ajello}, {Axelsson}, {Baldini}, {Ballet}, {Barbiellini}, {Bastieri}, {Baughman}, {Bechtol}, {Bellazzini}, {Berenji}, {Blandford}, {Bloom}, {Bonamente}, {Borgland}, {Brez}, {Brigida}, {Bruel}, {Burnett}, {Buson}, {Caliandro}, {Cameron}, {Caraveo}, {Casandjian}, {Cecchi}, {{\c{C}}elik}, {Chaty}, {Cheung}, {Chiang}, {Ciprini}, {Claus}, {Cohen-Tanugi}, {Cominsky}, {Conrad}, {Corbel}, {Corbet}, {Dermer}, {de Palma}, {Digel}, {do Couto e Silva}, {Drell}, {Dubois}, {Dubus}, {Dumora}, {Farnier}, {Favuzzi}, {Fegan}, {Focke}, {Fortin}, {Frailis}, {Fusco}, {Gargano}, {Gehrels}, {Germani}, {Giavitto}, {Giebels}, {Giglietto}, {Giordano}, {Glanzman}, {Godfrey}, {Grenier}, {Grondin}, {Grove}, {Guillemot}, {Guiriec}, {Hanabata}, {Harding}, {Hayashida}, {Hays}, {Hill}, {Hjalmarsdotter}, {Horan}, {Hughes}, {Jackson}, {J{\'o}hannesson}, {Johnson}, {Johnson}, {Johnson}, {Kamae}, {Katagiri}, {Kawai}, {Kerr}, {Kn{\"o}dlseder}, {Kocian}, {Koerding}, {Kuss}, {Lande},
  {Latronico}, {Lemoine-Goumard}, {Longo}, {Loparco}, {Lott}, {Lovellette}, {Lubrano}, {Madejski}, {Makeev}, {Marchand}, {Marelli}, {Max-Moerbeck}, {Mazziotta}, {McColl}, {McEnery}, {Meurer}, {Michelson}, {Migliari}, {Mitthumsiri}, {Mizuno}, {Monte}, {Monzani}, {Morselli}, {Moskalenko}, {Murgia}, {Nolan}, {Norris}, {Nuss}, {Ohsugi}, {Omodei}, {Ong}, {Ormes}, {Paneque}, {Parent}, {Pelassa}, {Pepe}, {Pesce-Rollins}, {Piron}, {Pooley}, {Porter}, {Pottschmidt}, {Rain{\`o}}, {Rando}, {Ray}, {Razzano}, {Rea}, {Readhead}, {Reimer}, {Reimer}, {Richards}, {Rochester}, {Rodriguez}, {Rodriguez}, {Romani}, {Ryde}, {Sadrozinski}, {Sander}, {Saz Parkinson}, {Sgr{\`o}}, {Siskind}, {Smith}, {Smith}, {Spinelli}, {Starck}, {Stevenson}, {Strickman}, {Suson}, {Takahashi}, {Tanaka}, {Thayer}, {Thompson}, {Tibaldo}, {Tomsick}, {Torres}, {Tosti}, {Tramacere}, {Uchiyama}, {Usher}, {Vasileiou}, {Vilchez}, {Vitale}, {Waite}, {Wang}, {Wilms}, {Winer}, {Wood}, {Ylinen}, \& {Ziegler}}]{Abdo2009}
{Abdo}, A.~A., {Ackermann}, M., {Ajello}, M., {et~al.} 2009, \bibinfo{title}{{Modulated High-Energy Gamma-Ray Emission from the Microquasar Cygnus X-3},} Science, 326, 1512, \dodoi{10.1126/science.1182174}

% type= article
\bibitem[{I.~I. {Antokhin} \& A.~M. {Cherepashchuk}(2019){Antokhin} \& {Cherepashchuk}}]{Antokhin2019}
{Antokhin}, I.~I., \& {Cherepashchuk}, A.~M. 2019, \bibinfo{title}{{The Period Change of Cyg X-3},} \apj, 871, 244, \dodoi{10.3847/1538-4357/aafb38}

% type= article
\bibitem[{I.~I. {Antokhin} {et~al.}(2022){Antokhin}, {Cherepashchuk}, {Antokhina}, \& {Tatarnikov}}]{Antokhin2022}
{Antokhin}, I.~I., {Cherepashchuk}, A.~M., {Antokhina}, E.~A., \& {Tatarnikov}, A.~M. 2022, \bibinfo{title}{{Near-IR and X-Ray Variability of Cyg X-3: Evidence for a Compact IR Source and Complex Wind Structures},} \apj, 926, 123, \dodoi{10.3847/1538-4357/ac4047}

% type= article
\bibitem[{W.~B. {Atwood} {et~al.}(2009){Atwood}, {Abdo}, {Ackermann}, {Althouse}, {Anderson}, {Axelsson}, {Baldini}, {Ballet}, {Band}, {Barbiellini}, {Bartelt}, {Bastieri}, {Baughman}, {Bechtol}, {B{\'e}d{\'e}r{\`e}de}, {Bellardi}, {Bellazzini}, {Berenji}, {Bignami}, {Bisello}, {Bissaldi}, {Blandford}, {Bloom}, {Bogart}, {Bonamente}, {Bonnell}, {Borgland}, {Bouvier}, {Bregeon}, {Brez}, {Brigida}, {Bruel}, {Burnett}, {Busetto}, {Caliandro}, {Cameron}, {Caraveo}, {Carius}, {Carlson}, {Casandjian}, {Cavazzuti}, {Ceccanti}, {Cecchi}, {Charles}, {Chekhtman}, {Cheung}, {Chiang}, {Chipaux}, {Cillis}, {Ciprini}, {Claus}, {Cohen-Tanugi}, {Condamoor}, {Conrad}, {Corbet}, {Corucci}, {Costamante}, {Cutini}, {Davis}, {Decotigny}, {DeKlotz}, {Dermer}, {de Angelis}, {Digel}, {do Couto e Silva}, {Drell}, {Dubois}, {Dumora}, {Edmonds}, {Fabiani}, {Farnier}, {Favuzzi}, {Flath}, {Fleury}, {Focke}, {Funk}, {Fusco}, {Gargano}, {Gasparrini}, {Gehrels}, {Gentit}, {Germani}, {Giebels}, {Giglietto}, {Giommi}, {Giordano}, {Glanzman},
  {Godfrey}, {Grenier}, {Grondin}, {Grove}, {Guillemot}, {Guiriec}, {Haller}, {Harding}, {Hart}, {Hays}, {Healey}, {Hirayama}, {Hjalmarsdotter}, {Horn}, {Hughes}, {J{\'o}hannesson}, {Johansson}, {Johnson}, {Johnson}, {Johnson}, {Johnson}, {Kamae}, {Katagiri}, {Kataoka}, {Kavelaars}, {Kawai}, {Kelly}, {Kerr}, {Klamra}, {Kn{\"o}dlseder}, {Kocian}, {Komin}, {Kuehn}, {Kuss}, {Landriu}, {Latronico}, {Lee}, {Lee}, {Lemoine-Goumard}, {Lionetto}, {Longo}, {Loparco}, {Lott}, {Lovellette}, {Lubrano}, {Madejski}, {Makeev}, {Marangelli}, {Massai}, {Mazziotta}, {McEnery}, {Menon}, {Meurer}, {Michelson}, {Minuti}, {Mirizzi}, {Mitthumsiri}, {Mizuno}, {Moiseev}, {Monte}, {Monzani}, {Moretti}, {Morselli}, {Moskalenko}, {Murgia}, {Nakamori}, {Nishino}, {Nolan}, {Norris}, {Nuss}, {Ohno}, {Ohsugi}, {Omodei}, {Orlando}, {Ormes}, {Paccagnella}, {Paneque}, {Panetta}, {Parent}, {Pearce}, {Pepe}, {Perazzo}, {Pesce-Rollins}, {Picozza}, {Pieri}, {Pinchera}, {Piron}, {Porter}, {Poupard}, {Rain{\`o}}, {Rando}, {Rapposelli}, {Razzano},
  {Reimer}, {Reimer}, {Reposeur}, {Reyes}, {Ritz}, {Rochester}, {Rodriguez}, {Romani}, {Roth}, {Russell}, {Ryde}, {Sabatini}, {Sadrozinski}, {Sanchez}, {Sander}, {Sapozhnikov}, {Parkinson}, {Scargle}, {Schalk}, \& {Scolieri}}]{Atwood2009}
{Atwood}, W.~B., {Abdo}, A.~A., {Ackermann}, M., {et~al.} 2009, \bibinfo{title}{{The Large Area Telescope on the Fermi Gamma-Ray Space Telescope Mission},} \apj, 697, 1071, \dodoi{10.1088/0004-637X/697/2/1071}

% type= article
\bibitem[{M. {Audard} {et~al.}(2024){Audard}, {Awaki}, {Ballhausen}, {Bamba}, {Behar}, {Boissay-Malaquin}, {Brenneman}, {Brown}, {Corrales}, {Costantini}, {Cumbee}, {D{\'\i}az Trigo}, {Done}, {Dotani}, {Ebisawa}, {Eckart}, {Eckert}, {Eguchi}, {Enoto}, {Ezoe}, {Foster}, {Fujimoto}, {Fujita}, {Fukazawa}, {Fukushima}, {Furuzawa}, {Gallo}, {Garc{\'\i}a}, {Gu}, {Guainazzi}, {Hagino}, {Hamaguchi}, {Hatsukade}, {Hayashi}, {Hayashi}, {Hell}, {Hodges-Kluck}, {Hornschemeier}, {Ichinohe}, {Ishida}, {Ishikawa}, {Ishisaki}, {Kaastra}, {Kallman}, {Kara}, {Katsuda}, {Kanemaru}, {Kelley}, {Kilbourne}, {Kitamoto}, {Kobayashi}, {Kohmura}, {Kubota}, {Leutenegger}, {Loewenstein}, {Maeda}, {Markevitch}, {Matsumoto}, {Matsushita}, {McCammon}, {McNamara}, {Mernier}, {Miller}, {Miller}, {Mitsuishi}, {Mizumoto}, {Mizuno}, {Mori}, {Mukai}, {Murakami}, {Mushotzky}, {Nakajima}, {Nakazawa}, {Ness}, {Nobukawa}, {Nobukawa}, {Noda}, {Odaka}, {Ogawa}, {Ogorzalek}, {Okajima}, {Ota}, {Paltani}, {Petre}, {Plucinsky}, {Porter}, {Pottschmidt},
  {Sato}, {Sato}, {Sawada}, {Seta}, {Shidatsu}, {Simionescu}, {Smith}, {Suzuki}, {Szymkowiak}, {Takahashi}, {Takeo}, {Tamagawa}, {Tamura}, {Tanaka}, {Tanimoto}, {Tashiro}, {Terada}, {Terashima}, {Tsuboi}, {Tsujimoto}, {Tsunemi}, {Tsuru}, {Uchida}, {Uchida}, {Uchida}, {Uchiyama}, {Ueda}, {Uno}, {Vink}, {Watanabe}, {Williams}, {Yamada}, {Yamada}, {Yamaguchi}, {Yamaoka}, {Yamasaki}, {Yamauchi}, {Yamauchi}, {Yaqoob}, {Yoneyama}, {Yoshida}, {Yukita}, {Zhuravleva}, {Tomaru}, {Hayashi}, {Hakamata}, {Miura}, {Koljonen}, \& {McCollough}}]{Audard2024}
{Audard}, M., {Awaki}, H., {Ballhausen}, R., {et~al.} 2024, \bibinfo{title}{{The XRISM/Resolve View of the Fe K Region of Cyg X-3},} \apjl, 977, L34, \dodoi{10.3847/2041-8213/ad8ed0}

% type= article
\bibitem[{L. {Baldini} {et~al.}(2022){Baldini}, {Bucciantini}, {Lalla}, {Ehlert}, {Manfreda}, {Negro}, {Omodei}, {Pesce-Rollins}, {Sgr{\`o}}, \& {Silvestri}}]{Baldini2022}
{Baldini}, L., {Bucciantini}, N., {Lalla}, N.~D., {et~al.} 2022, \bibinfo{title}{{ixpeobssim: A simulation and analysis framework for the imaging X-ray polarimetry explorer},} SoftwareX, 19, 101194, \dodoi{10.1016/j.softx.2022.101194}

% type= article
\bibitem[{A.~M. {Beloborodov}(1998){Beloborodov}}]{Beloborodov1998}
{Beloborodov}, A.~M. 1998, \bibinfo{title}{{Polarization Change Due to Fast Winds from Accretion Disks},} \apjl, 496, L105, \dodoi{10.1086/311260}

% type= article
\bibitem[{B. {Cerutti} {et~al.}(2011){Cerutti}, {Dubus}, {Malzac}, {Szostek}, {Belmont}, {Zdziarski}, \& {Henri}}]{Cerutti2011}
{Cerutti}, B., {Dubus}, G., {Malzac}, J., {et~al.} 2011, \bibinfo{title}{{Absorption of high-energy gamma rays in Cygnus X-3},} \aap, 529, A120, \dodoi{10.1051/0004-6361/201116581}

% type= article
\bibitem[{S. {Corbel} {et~al.}(2012){Corbel}, {Dubus}, {Tomsick}, {Szostek}, {Corbet}, {Miller-Jones}, {Richards}, {Pooley}, {Trushkin}, {Dubois}, {Hill}, {Kerr}, {Max-Moerbeck}, {Readhead}, {Bodaghee}, {Tudose}, {Parent}, {Wilms}, \& {Pottschmidt}}]{Corbel2012}
{Corbel}, S., {Dubus}, G., {Tomsick}, J.~A., {et~al.} 2012, \bibinfo{title}{{A giant radio flare from Cygnus X-3 with associated {\ensuremath{\gamma}}-ray emission},} \mnras, 421, 2947, \dodoi{10.1111/j.1365-2966.2012.20517.x}

% type= article
\bibitem[{A. {Di Marco} {et~al.}(2022){Di Marco}, {Costa}, {Muleri}, {Soffitta}, {Fabiani}, {La Monaca}, {Rankin}, {Xie}, {Bachetti}, {Baldini}, {Baumgartner}, {Bellazzini}, {Brez}, {Castellano}, {Del Monte}, {Di Lalla}, {Ferrazzoli}, {Latronico}, {Maldera}, {Manfreda}, {O'Dell}, {Perri}, {Pesce-Rollins}, {Puccetti}, {Ramsey}, {Ratheesh}, {Sgr{\`o}}, {Spandre}, {Tennant}, {Tobia}, {Trois}, \& {Weisskopf}}]{DiMarco2022}
{Di Marco}, A., {Costa}, E., {Muleri}, F., {et~al.} 2022, \bibinfo{title}{{A Weighted Analysis to Improve the X-Ray Polarization Sensitivity of the Imaging X-ray Polarimetry Explorer},} \aj, 163, 170, \dodoi{10.3847/1538-3881/ac51c9}

% type= article
\bibitem[{A. {Di Marco et al.}(2023){Di Marco et al.}}]{DiMarco2023}
{Di Marco et al.}, A. 2023, \bibinfo{title}{{Handling the Background in IXPE Polarimetric Data},} \aj, 165, 143, \dodoi{10.3847/1538-3881/acba0f}

% type= article
\bibitem[{M. {Dov{\v{c}}iak} {et~al.}(2024){Dov{\v{c}}iak}, {Podgorn{\'y}}, {Svoboda}, {Steiner}, {Kaaret}, {Krawczynski}, {Ingram}, {Kravtsov}, {Marra}, {Muleri}, {Garc{\'\i}a}, {Mastroserio}, {Miku{\v{s}}incov{\'a}}, {Ratheesh}, \& {Cavero}}]{Dovciak2024}
{Dov{\v{c}}iak}, M., {Podgorn{\'y}}, J., {Svoboda}, J., {et~al.} 2024, \bibinfo{title}{{IXPE View of BH XRBs during the First 2.5 Years of the Mission},} Galaxies, 12, 54, \dodoi{10.3390/galaxies12050054}

% type= article
\bibitem[{G. {Dubus} {et~al.}(2010){Dubus}, {Cerutti}, \& {Henri}}]{Dubus2010}
{Dubus}, G., {Cerutti}, B., \& {Henri}, G. 2010, \bibinfo{title}{{The relativistic jet of Cygnus X-3 in gamma-rays},} \mnras, 404, L55, \dodoi{10.1111/j.1745-3933.2010.00834.x}

% type= article
\bibitem[{M. {Ewing} {et~al.}(2025){Ewing}, {Parra}, {Mastroserio}, {Veledina}, {Ingram}, {Dov{\v{c}}iak}, {Garc{\'\i}a}, {Russell}, {Baglio}, {Poutanen}, {Adegoke}, {Bianchi}, {Capitanio}, {Connors}, {Del Santo}, {De Marco}, {Trigo}, {Gandhi}, {Gupta}, {Kang}, {Kammoun}, {Loktev}, {Marra}, {Matt}, {Nathan}, {Petrucci}, {Shidatsu}, {Steiner}, {Tombesi}, \& {Vincentelli}}]{Ewing2025}
{Ewing}, M., {Parra}, M., {Mastroserio}, G., {et~al.} 2025, \bibinfo{title}{{The very high X-ray polarization of accreting black hole IGR J17091‑3624 in the hard state},} \mnras, 541, 1774, \dodoi{10.1093/mnras/staf859}

% type= article
\bibitem[{R. {Giacconi} {et~al.}(1967){Giacconi}, {Gorenstein}, {Gursky}, \& {Waters}}]{Giacconi1967}
{Giacconi}, R., {Gorenstein}, P., {Gursky}, H., \& {Waters}, J.~R. 1967, \bibinfo{title}{{An X-Ray Survey of the Cygnus Region},} \apjl, 148, L119, \dodoi{10.1086/180028}

% type= article
\bibitem[{P.~C. {Gregory} {et~al.}(1972){Gregory}, {Kronberg}, {Seaquist}, {Hughes}, {Woodsworth}, {Viner}, \& {Retallack}}]{Gregory1972}
{Gregory}, P.~C., {Kronberg}, P.~P., {Seaquist}, E.~R., {et~al.} 1972, \bibinfo{title}{{Discovery of Giant Radio Outburst from Cygnus X-3},} \nat, 239, 440, \dodoi{10.1038/239440a0}

% type= article
\bibitem[{L. {Hjalmarsdotter} {et~al.}(2008){Hjalmarsdotter}, {Zdziarski}, {Larsson}, {Beckmann}, {McCollough}, {Hannikainen}, \& {Vilhu}}]{Hjalmarsdotter2008}
{Hjalmarsdotter}, L., {Zdziarski}, A.~A., {Larsson}, S., {et~al.} 2008, \bibinfo{title}{{The nature of the hard state of Cygnus X-3},} \mnras, 384, 278, \dodoi{10.1111/j.1365-2966.2007.12688.x}

% type= article
\bibitem[{L. {Hjalmarsdotter} {et~al.}(2009){Hjalmarsdotter}, {Zdziarski}, {Szostek}, \& {Hannikainen}}]{Hjalmarsdotter2009}
{Hjalmarsdotter}, L., {Zdziarski}, A.~A., {Szostek}, A., \& {Hannikainen}, D.~C. 2009, \bibinfo{title}{{Spectral variability in Cygnus X-3},} \mnras, 392, 251, \dodoi{10.1111/j.1365-2966.2008.14036.x}

% type= article
\bibitem[{A. {Ingram} {et~al.}(2024){Ingram}, {Bollemeijer}, {Veledina}, {Dov{\v{c}}iak}, {Poutanen}, {Egron}, {Russell}, {Trushkin}, {Negro}, {Ratheesh}, {Capitanio}, {Connors}, {Neilsen}, {Kraus}, {Iacolina}, {Pellizzoni}, {Pilia}, {Carotenuto}, {Matt}, {Mastroserio}, {Kaaret}, {Bianchi}, {Garc{\'\i}a}, {Bachetti}, {Wu}, {Costa}, {Ewing}, {Kravtsov}, {Krawczynski}, {Loktev}, {Marinucci}, {Marra}, {Miku{\v{s}}incov{\'a}}, {Nathan}, {Parra}, {Petrucci}, {Righini}, {Soffitta}, {Steiner}, {Svoboda}, {Tombesi}, {Tugliani}, {Ursini}, {Yang}, {Zane}, {Zhang}, {Agudo}, {Antonelli}, {Baldini}, {Baumgartner}, {Bellazzini}, {Bongiorno}, {Bonino}, {Brez}, {Bucciantini}, {Castellano}, {Cavazzuti}, {Chen}, {Ciprini}, {De Rosa}, {Del Monte}, {Di Gesu}, {Di Lalla}, {Di Marco}, {Donnarumma}, {Doroshenko}, {Ehlert}, {Enoto}, {Evangelista}, {Fabiani}, {Ferrazzoli}, {Gunji}, {Hayashida}, {Heyl}, {Iwakiri}, {Jorstad}, {Karas}, {Kislat}, {Kitaguchi}, {Kolodziejczak}, {La Monaca}, {Latronico}, {Liodakis}, {Maldera}, {Manfreda},
  {Marin}, {Marscher}, {Marshall}, {Massaro}, {Mitsuishi}, {Mizuno}, {Muleri}, {Ng}, {O'Dell}, {Omodei}, {Oppedisano}, {Papitto}, {Pavlov}, {Peirson}, {Perri}, {Pesce-Rollins}, {Possenti}, {Puccetti}, {Ramsey}, {Rankin}, {Roberts}, {Romani}, {Sgr{\`o}}, {Slane}, {Spandre}, {Swartz}, {Tamagawa}, {Tavecchio}, {Taverna}, {Tawara}, {Tennant}, {Thomas}, {Trois}, {Tsygankov}, {Turolla}, {Vink}, {Weisskopf}, {Xie}, \& {IXPE Collaboration}}]{Ingram2024}
{Ingram}, A., {Bollemeijer}, N., {Veledina}, A., {et~al.} 2024, \bibinfo{title}{{Tracking the X-Ray Polarization of the Black Hole Transient Swift J1727.8{\textendash}1613 during a State Transition},} \apj, 968, 76, \dodoi{10.3847/1538-4357/ad3faf}

% type= article
\bibitem[{T. {Kallman} {et~al.}(2019){Kallman}, {McCollough}, {Koljonen}, {Liedahl}, {Miller}, {Paerels}, {Pooley}, {Sako}, {Schulz}, {Trushkin}, \& {Corrales}}]{Kallman2019}
{Kallman}, T., {McCollough}, M., {Koljonen}, K., {et~al.} 2019, \bibinfo{title}{{Photoionization Emission Models for the Cyg X-3 X-Ray Spectrum},} \apj, 874, 51, \dodoi{10.3847/1538-4357/ab09f8}

% type= article
\bibitem[{K.~I.~I. {Koljonen} {et~al.}(2023){Koljonen}, {Satalecka}, {Lindfors}, \& {Liodakis}}]{Koljonen2023}
{Koljonen}, K. I.~I., {Satalecka}, K., {Lindfors}, E.~J., \& {Liodakis}, I. 2023, \bibinfo{title}{{Microquasar Cyg X-3 - a unique jet-wind neutrino factory?},} \mnras, 524, L89, \dodoi{10.1093/mnrasl/slad081}

% type= article
\bibitem[{V. {Kravtsov} {et~al.}(2025){Kravtsov}, {Bocharova}, {Veledina}, {Poutanen}, {Hughes}, {Dov{\v{c}}iak}, {Egron}, {Muleri}, {Podgorny}, {Svoboda}, {Forsblom}, {Berdyugin}, {Blinov}, {Bright}, {Carotenuto}, {Green}, {Ingram}, {Liodakis}, {Mandarakas}, {Nitindala}, {Rhodes}, {Trushkin}, {Tsygankov}, {Brigitte}, {Di Marco}, {Iacolina}, {Krawczynski}, {La Monaca}, {Loktev}, {Mastroserio}, {Petrucci}, {Pilia}, {Tombesi}, \& {Zdziarski}}]{Kravtsov2025}
{Kravtsov}, V., {Bocharova}, A., {Veledina}, A., {et~al.} 2025, \bibinfo{title}{{Variability of X-ray polarization of Cyg X-1},} \aap, in press, arXiv:2505.03942, \dodoi{10.1051/0004-6361/202555411}

% type= article
\bibitem[{H. {Krawczynski} \& K. {Hu}(2025){Krawczynski} \& {Hu}}]{Krawczynski2025}
{Krawczynski}, H., \& {Hu}, K. 2025, \bibinfo{title}{{The Cygnus X-1 Puzzle: Implications of X-ray Polarization Measurements in the Soft and Hard States on the Properties of the Accretion Flow and the Emission Mechanisms},} arXiv e-prints, arXiv:2506.01184, \dodoi{10.48550/arXiv.2506.01184}

% type= article
\bibitem[{H. {Krawczynski} {et~al.}(2022){Krawczynski}, {Muleri}, {Dov{\v{c}}iak}, {Veledina}, {Rodriguez Cavero}, {Svoboda}, {Ingram}, {Matt}, {Garcia}, {Loktev}, {Negro}, {Poutanen}, {Kitaguchi}, {Podgorn{\'y}}, {Rankin}, {Zhang}, {Berdyugin}, {Berdyugina}, {Bianchi}, {Blinov}, {Capitanio}, {Di Lalla}, {Draghis}, {Fabiani}, {Kagitani}, {Kravtsov}, {Kiehlmann}, {Latronico}, {Lutovinov}, {Mandarakas}, {Marin}, {Marinucci}, {Miller}, {Mizuno}, {Molkov}, {Omodei}, {Petrucci}, {Ratheesh}, {Sakanoi}, {Semena}, {Skalidis}, {Soffitta}, {Tennant}, {Thalhammer}, {Tombesi}, {Weisskopf}, {Wilms}, {Zhang}, {Agudo}, {Antonelli}, {Bachetti}, {Baldini}, {Baumgartner}, {Bellazzini}, {Bongiorno}, {Bonino}, {Brez}, {Bucciantini}, {Castellano}, {Cavazzuti}, {Ciprini}, {Costa}, {De Rosa}, {Del Monte}, {Di Gesu}, {Di Marco}, {Donnarumma}, {Doroshenko}, {Ehlert}, {Enoto}, {Evangelista}, {Ferrazzoli}, {Gunji}, {Hayashida}, {Heyl}, {Iwakiri}, {Jorstad}, {Karas}, {Kolodziejczak}, {La Monaca}, {Liodakis}, {Maldera}, {Manfreda},
  {Marscher}, {Marshall}, {Mitsuishi}, {Ng}, {O{\textquoteright}Dell}, {Oppedisano}, {Papitto}, {Pavlov}, {Peirson}, {Perri}, {Pesce-Rollins}, {Pilia}, {Possenti}, {Puccetti}, {Ramsey}, {Romani}, {Sgr{\`o}}, {Slane}, {Spandre}, {Tamagawa}, {Tavecchio}, {Taverna}, {Tawara}, {Thomas}, {Trois}, {Tsygankov}, {Turolla}, {Vink}, {Wu}, {Xie}, \& {Zane}}]{Krawczynski2022}
{Krawczynski}, H., {Muleri}, F., {Dov{\v{c}}iak}, M., {et~al.} 2022, \bibinfo{title}{{Polarized x-rays constrain the disk-jet geometry in the black hole x-ray binary Cygnus X-1},} Science, 378, 650, \dodoi{10.1126/science.add5399}

% type= article
\bibitem[{M.~L. {McCollough} {et~al.}(1999){McCollough}, {Robinson}, {Zhang}, {Harmon}, {Hjellming}, {Waltman}, {Foster}, {Ghigo}, {Briggs}, {Pendleton}, \& {Johnston}}]{McCollough1999}
{McCollough}, M.~L., {Robinson}, C.~R., {Zhang}, S.~N., {et~al.} 1999, \bibinfo{title}{{Discovery of Correlated Behavior between the Hard X-Ray and the Radio Bands in Cygnus X-3},} \apj, 517, 951, \dodoi{10.1086/307241}

% type= article
\bibitem[{D.~R. {Parsignault} {et~al.}(1972){Parsignault}, {Gursky}, {Kellogg}, {Matilsky}, {Murray}, {Schreier}, {Tananbaum}, {Giacconi}, \& {Brinkman}}]{Parsignault1972}
{Parsignault}, D.~R., {Gursky}, H., {Kellogg}, E.~M., {et~al.} 1972, \bibinfo{title}{{Observations of Cygnus X-3 by Uhuru},} Nature Physical Science, 239, 123, \dodoi{10.1038/physci239123a0}

% type= article
\bibitem[{G. {Piano} {et~al.}(2012){Piano}, {Tavani}, {Vittorini}, {Trois}, {Giuliani}, {Bulgarelli}, {Evangelista}, {Coppi}, {Del Monte}, {Sabatini}, {Striani}, {Donnarumma}, {Hannikainen}, {Koljonen}, {McCollough}, {Pooley}, {Trushkin}, {Zanin}, {Barbiellini}, {Cardillo}, {Cattaneo}, {Chen}, {Colafrancesco}, {Feroci}, {Fuschino}, {Giusti}, {Longo}, {Morselli}, {Pellizzoni}, {Pittori}, {Pucella}, {Rapisarda}, {Rappoldi}, {Soffitta}, {Trifoglio}, {Vercellone}, \& {Verrecchia}}]{Piano2012}
{Piano}, G., {Tavani}, M., {Vittorini}, V., {et~al.} 2012, \bibinfo{title}{{The AGILE monitoring of Cygnus X-3: transient gamma-ray emission and spectral constraints},} \aap, 545, A110, \dodoi{10.1051/0004-6361/201219145}

% type= article
\bibitem[{J. {Podgorn{\'y}} {et~al.}(2024){Podgorn{\'y}}, {Svoboda}, {Dov{\v{c}}iak}, {Veledina}, {Poutanen}, {Kaaret}, {Bianchi}, {Ingram}, {Capitanio}, {Datta}, {Egron}, {Krawczynski}, {Matt}, {Muleri}, {Petrucci}, {Russell}, {Steiner}, {Bollemeijer}, {Brigitte}, {Castro Segura}, {Emami}, {Garc{\'\i}a}, {Hu}, {Iacolina}, {Kravtsov}, {Marra}, {Mastroserio}, {Mu{\~n}oz-Darias}, {Nathan}, {Negro}, {Ratheesh}, {Rodriguez Cavero}, {Taverna}, {Tombesi}, {Yang}, {Zhang}, \& {Zhang}}]{Podgorny2024}
{Podgorn{\'y}}, J., {Svoboda}, J., {Dov{\v{c}}iak}, M., {et~al.} 2024, \bibinfo{title}{{Recovery of the X-ray polarisation of Swift J1727.8{\ensuremath{-}}1613 after the soft-to-hard spectral transition},} \aap, 686, L12, \dodoi{10.1051/0004-6361/202450566}

% type= article
\bibitem[{J. {Poutanen} {et~al.}(2023){Poutanen}, {Veledina}, \& {Beloborodov}}]{Poutanen2023}
{Poutanen}, J., {Veledina}, A., \& {Beloborodov}, A.~M. 2023, \bibinfo{title}{{Polarized X-Rays from Windy Accretion in Cygnus X-1},} \apjl, 949, L10, \dodoi{10.3847/2041-8213/acd33e}

% type= article
\bibitem[{M.~J. {Reid} \& J.~C.~A. {Miller-Jones}(2023){Reid} \& {Miller-Jones}}]{Reid2023}
{Reid}, M.~J., \& {Miller-Jones}, J.~C.~A. 2023, \bibinfo{title}{{On the Distances to the X-Ray Binaries Cygnus X-3 and GRS 1915+105},} \apj, 959, 85, \dodoi{10.3847/1538-4357/acfe0c}

% type= article
\bibitem[{G.~E. {Romero} {et~al.}(2003){Romero}, {Torres}, {Kaufman Bernad{\'o}}, \& {Mirabel}}]{Romero2003}
{Romero}, G.~E., {Torres}, D.~F., {Kaufman Bernad{\'o}}, M.~M., \& {Mirabel}, I.~F. 2003, \bibinfo{title}{{Hadronic gamma-ray emission from windy microquasars},} \aap, 410, L1, \dodoi{10.1051/0004-6361:20031314-1}

% type= article
\bibitem[{P. {Soffitta} {et~al.}(2021){Soffitta}, {Baldini}, {Bellazzini}, {Costa}, {Latronico}, {Muleri}, {Del Monte}, {Fabiani}, {Minuti}, {Pinchera}, {Sgro'}, {Spandre}, {Trois}, {Amici}, {Andersson}, {Attina'}, {Bachetti}, {Barbanera}, {Borotto}, {Brez}, {Brienza}, {Caporale}, {Cardelli}, {Carpentiero}, {Castellano}, {Castronuovo}, {Cavalli}, {Cavazzuti}, {Ceccanti}, {Centrone}, {Ciprini}, {Citraro}, {D'Amico}, {D'Alba}, {Di Cosimo}, {Di Lalla}, {Di Marco}, {Di Persio}, {Donnarumma}, {Evangelista}, {Ferrazzoli}, {Hayato}, {Kitaguchi}, {La Monaca}, {Lefevre}, {Loffredo}, {Lorenzi}, {Lucchesi}, {Magazzu}, {Maldera}, {Manfreda}, {Mangraviti}, {Marengo}, {Matt}, {Mereu}, {Morbidini}, {Mosti}, {Nakano}, {Nasimi}, {Negri}, {Nenonen}, {Nuti}, {Orsini}, {Perri}, {Pesce-Rollins}, {Piazzolla}, {Pilia}, {Profeti}, {Puccetti}, {Rankin}, {Ratheesh}, {Rubini}, {Santoli}, {Sarra}, {Scalise}, {Sciortino}, {Tamagawa}, {Tardiola}, {Tobia}, {Vimercati}, \& {Xie}}]{Soffitta2021}
{Soffitta}, P., {Baldini}, L., {Bellazzini}, R., {et~al.} 2021, \bibinfo{title}{{The Instrument of the Imaging X-Ray Polarimetry Explorer},} \aj, 162, 208, \dodoi{10.3847/1538-3881/ac19b0}

% type= article
\bibitem[{R.~E. {Spencer} {et~al.}(2022){Spencer}, {Garrett}, {Bray}, \& {Green}}]{Spencer2022}
{Spencer}, R.~E., {Garrett}, M., {Bray}, J.~D., \& {Green}, D.~A. 2022, \bibinfo{title}{{Major and minor flares on Cygnus X-3 revisited},} \mnras, 512, 2618, \dodoi{10.1093/mnras/stac666}

% type= article
\bibitem[{A. {Szostek} {et~al.}(2008){Szostek}, {Zdziarski}, \& {McCollough}}]{Szostek2008}
{Szostek}, A., {Zdziarski}, A.~A., \& {McCollough}, M.~L. 2008, \bibinfo{title}{{A classification of the X-ray and radio states of Cyg X-3 and their long-term correlations},} \mnras, 388, 1001, \dodoi{10.1111/j.1365-2966.2008.13479.x}

% type= article
\bibitem[{M. {Tavani} {et~al.}(2009){Tavani}, {Bulgarelli}, {Piano}, {Sabatini}, {Striani}, {Evangelista}, {Trois}, {Pooley}, {Trushkin}, {Nizhelskij}, {McCollough}, {Koljonen}, {Pucella}, {Giuliani}, {Chen}, {Costa}, {Vittorini}, {Trifoglio}, {Gianotti}, {Argan}, {Barbiellini}, {Caraveo}, {Cattaneo}, {Cocco}, {Contessi}, {D'Ammando}, {Del Monte}, {de Paris}, {Di Cocco}, {di Persio}, {Donnarumma}, {Feroci}, {Ferrari}, {Fuschino}, {Galli}, {Labanti}, {Lapshov}, {Lazzarotto}, {Lipari}, {Longo}, {Mattaini}, {Marisaldi}, {Mastropietro}, {Mauri}, {Mereghetti}, {Morelli}, {Morselli}, {Pacciani}, {Pellizzoni}, {Perotti}, {Picozza}, {Pilia}, {Prest}, {Rapisarda}, {Rappoldi}, {Rossi}, {Rubini}, {Scalise}, {Soffitta}, {Vallazza}, {Vercellone}, {Zambra}, {Zanello}, {Pittori}, {Verrecchia}, {Giommi}, {Colafrancesco}, {Santolamazza}, {Antonelli}, \& {Salotti}}]{Tavani2009}
{Tavani}, M., {Bulgarelli}, A., {Piano}, G., {et~al.} 2009, \bibinfo{title}{{Extreme particle acceleration in the microquasar CygnusX-3},} \nat, 462, 620, \dodoi{10.1038/nature08578}

% type= article
\bibitem[{S. {Trushkin} {et~al.}(2017){Trushkin}, {McCollough}, {Nizhelskij}, \& {Tsybulev}}]{Trushkin2017}
{Trushkin}, S., {McCollough}, M., {Nizhelskij}, N., \& {Tsybulev}, P. 2017, \bibinfo{title}{{The Giant Flares of the Microquasar Cygnus X-3: X-Rays States and Jets},} Galaxies, 5, 86, \dodoi{10.3390/galaxies5040086}

% type= article
\bibitem[{F. {Ursini} {et~al.}(2023){Ursini}, {Marinucci}, {Matt}, {Bianchi}, {Marin}, {Marshall}, {Middei}, {Poutanen}, {Rogantini}, {De Rosa}, {Di Gesu}, {Garc{\'\i}a}, {Ingram}, {Kim}, {Krawczynski}, {Puccetti}, {Soffitta}, {Svoboda}, {Tombesi}, {Weisskopf}, {Barnouin}, {Perri}, {Podgorny}, {Ratheesh}, {Zaino}, {Agudo}, {Antonelli}, {Bachetti}, {Baldini}, {Baumgartner}, {Bellazzini}, {Bongiorno}, {Bonino}, {Brez}, {Bucciantini}, {Capitanio}, {Castellano}, {Cavazzuti}, {Ciprini}, {Costa}, {Del Monte}, {Di Lalla}, {Di Marco}, {Donnarumma}, {Doroshenko}, {Dovciak}, {Ehlert}, {Enoto}, {Evangelista}, {Fabiani}, {Ferrazzoli}, {Gunji}, {Heyl}, {Iwakiri}, {Jorstad}, {Karas}, {Kitaguchi}, {Kolodziejczak}, {La Monaca}, {Latronico}, {Liodakis}, {Maldera}, {Manfreda}, {Marscher}, {Mitsuishi}, {Mizuno}, {Muleri}, {Ng}, {O'Dell}, {Omodei}, {Oppedisano}, {Papitto}, {Pavlov}, {Peirson}, {Pesce-Rollins}, {Petrucci}, {Pilia}, {Possenti}, {Ramsey}, {Rankin}, {Romani}, {Sgr{\`o}}, {Slane}, {Spandre}, {Tamagawa}, {Tavecchio},
  {Taverna}, {Tawara}, {Tennant}, {Thomas}, {Trois}, {Tsygankov}, {Turolla}, {Vink}, {Wu}, {Xie}, \& {Zane}}]{Ursini2023}
{Ursini}, F., {Marinucci}, A., {Matt}, G., {et~al.} 2023, \bibinfo{title}{{Mapping the circumnuclear regions of the Circinus galaxy with the Imaging X-ray Polarimetry Explorer},} \mnras, 519, 50, \dodoi{10.1093/mnras/stac3189}

% type= article
\bibitem[{M.~H. {van Kerkwijk} {et~al.}(1996){van Kerkwijk}, {Geballe}, {King}, {van der Klis}, \& {van Paradijs}}]{Kerkwijk1996}
{van Kerkwijk}, M.~H., {Geballe}, T.~R., {King}, D.~L., {van der Klis}, M., \& {van Paradijs}, J. 1996, \bibinfo{title}{{The Wolf-Rayet counterpart of Cygnus X-3.},} \aap, 314, 521, \dodoi{10.48550/arXiv.astro-ph/9604100}

% type= article
\bibitem[{A. {Veledina} {et~al.}(2023){Veledina}, {Muleri}, {Dov{\v{c}}iak}, {Poutanen}, {Ratheesh}, {Capitanio}, {Matt}, {Soffitta}, {Tennant}, {Negro}, {Kaaret}, {Costa}, {Ingram}, {Svoboda}, {Krawczynski}, {Bianchi}, {Steiner}, {Garc{\'\i}a}, {Kravtsov}, {Nitindala}, {Ewing}, {Mastroserio}, {Marinucci}, {Ursini}, {Tombesi}, {Tsygankov}, {Yang}, {Weisskopf}, {Trushkin}, {Egron}, {Iacolina}, {Pilia}, {Marra}, {Miku{\v{s}}incov{\'a}}, {Nathan}, {Parra}, {Petrucci}, {Podgorn{\'y}}, {Tugliani}, {Zane}, {Zhang}, {Agudo}, {Antonelli}, {Bachetti}, {Baldini}, {Baumgartner}, {Bellazzini}, {Bongiorno}, {Bonino}, {Brez}, {Bucciantini}, {Castellano}, {Cavazzuti}, {Chen}, {Ciprini}, {De Rosa}, {Del Monte}, {Di Gesu}, {Di Lalla}, {Di Marco}, {Donnarumma}, {Doroshenko}, {Ehlert}, {Enoto}, {Evangelista}, {Fabiani}, {Ferrazzoli}, {Gunji}, {Hayashida}, {Heyl}, {Iwakiri}, {Jorstad}, {Karas}, {Kislat}, {Kitaguchi}, {Kolodziejczak}, {La Monaca}, {Latronico}, {Liodakis}, {Maldera}, {Manfreda}, {Marin}, {Marscher}, {Marshall},
  {Massaro}, {Mitsuishi}, {Mizuno}, {Ng}, {O'Dell}, {Omodei}, {Oppedisano}, {Papitto}, {Pavlov}, {Peirson}, {Perri}, {Pesce-Rollins}, {Possenti}, {Puccetti}, {Ramsey}, {Rankin}, {Roberts}, {Romani}, {Sgr{\`o}}, {Slane}, {Spandre}, {Swartz}, {Tamagawa}, {Tavecchio}, {Taverna}, {Tawara}, {Thomas}, {Trois}, {Turolla}, {Vink}, {Wu}, \& {Xie}}]{Veledina2023}
{Veledina}, A., {Muleri}, F., {Dov{\v{c}}iak}, M., {et~al.} 2023, \bibinfo{title}{{Discovery of X-Ray Polarization from the Black Hole Transient Swift J1727.8-1613},} \apjl, 958, L16, \dodoi{10.3847/2041-8213/ad0781}

% type= article
\bibitem[{A. {Veledina} {et~al.}(2024{\natexlab{a}}){Veledina}, {Muleri}, {Poutanen}, {Podgorn{\'y}}, {Dov{\v{c}}iak}, {Capitanio}, {Churazov}, {De Rosa}, {Di Marco}, {Forsblom}, {Kaaret}, {Krawczynski}, {La Monaca}, {Loktev}, {Lutovinov}, {Molkov}, {Mushtukov}, {Ratheesh}, {Rodriguez Cavero}, {Steiner}, {Sunyaev}, {Tsygankov}, {Weisskopf}, {Zdziarski}, {Bianchi}, {Bright}, {Bursov}, {Costa}, {Egron}, {Garcia}, {Green}, {Gurwell}, {Ingram}, {Kajava}, {Kale}, {Kraus}, {Malyshev}, {Marin}, {Matt}, {McCollough}, {Mereminskiy}, {Nizhelsky}, {Piano}, {Pilia}, {Pittori}, {Rao}, {Righini}, {Soffitta}, {Shevchenko}, {Svoboda}, {Tombesi}, {Trushkin}, {Tsybulev}, {Ursini}, {Wu}, {Agudo}, {Antonelli}, {Bachetti}, {Baldini}, {Baumgartner}, {Bellazzini}, {Bongiorno}, {Bonino}, {Brez}, {Bucciantini}, {Castellano}, {Cavazzuti}, {Chen}, {Ciprini}, {Del Monte}, {Di Gesu}, {Di Lalla}, {Donnarumma}, {Doroshenko}, {Ehlert}, {Enoto}, {Evangelista}, {Fabiani}, {Ferrazzoli}, {Gunji}, {Hayashida}, {Heyl}, {Iwakiri}, {Jorstad},
  {Karas}, {Kislat}, {Kitaguchi}, {Kolodziejczak}, {Latronico}, {Liodakis}, {Maldera}, {Manfreda}, {Marinucci}, {Marscher}, {Marshall}, {Massaro}, {Mitsuishi}, {Mizuno}, {Negro}, {Ng}, {O'Dell}, {Omodei}, {Oppedisano}, {Papitto}, {Pavlov}, {Peirson}, {Perri}, {Pesce-Rollins}, {Petrucci}, {Possenti}, {Puccetti}, {Ramsey}, {Rankin}, {Roberts}, {Romani}, {Sgr{\`o}}, {Slane}, {Spandre}, {Swartz}, {Tamagawa}, {Tavecchio}, {Taverna}, {Tawara}, {Tennant}, {Thomas}, {Trois}, {Turolla}, {Vink}, {Xie}, \& {Zane}}]{Veledina2024}
{Veledina}, A., {Muleri}, F., {Poutanen}, J., {et~al.} 2024{\natexlab{a}}, \bibinfo{title}{{Cygnus X-3 revealed as a Galactic ultraluminous X-ray source by IXPE},} Nature Astronomy, 8, 1031, \dodoi{10.1038/s41550-024-02294-9}

% type= article
\bibitem[{A. {Veledina} {et~al.}(2024{\natexlab{b}}){Veledina}, {Poutanen}, {Bocharova}, {Di Marco}, {Forsblom}, {La Monaca}, {Podgorn{\'y}}, {Tsygankov}, {Zdziarski}, {Ahlberg}, {Green}, {Muleri}, {Rhodes}, {Bianchi}, {Costa}, {Dov{\v{c}}iak}, {Loktev}, {McCollough}, {Soffitta}, \& {Sunyaev}}]{Veledina2024b}
{Veledina}, A., {Poutanen}, J., {Bocharova}, A., {et~al.} 2024{\natexlab{b}}, \bibinfo{title}{{Ultrasoft state of microquasar Cygnus X-3: X-ray polarimetry reveals the geometry of the astronomical puzzle},} \aap, 688, L27, \dodoi{10.1051/0004-6361/202451356}

% type= article
\bibitem[{M.~C. {Weisskopf} {et~al.}(2022){Weisskopf}, {Soffitta}, {Baldini}, {Ramsey}, {O'Dell}, {Romani}, {Matt}, {Deininger}, {Baumgartner}, {Bellazzini}, {Costa}, {Kolodziejczak}, {Latronico}, {Marshall}, {Muleri}, {Bongiorno}, {Tennant}, {Bucciantini}, {Dovciak}, {Marin}, {Marscher}, {Poutanen}, {Slane}, {Turolla}, {Kalinowski}, {Di Marco}, {Fabiani}, {Minuti}, {La Monaca}, {Pinchera}, {Rankin}, {Sgro'}, {Trois}, {Xie}, {Alexander}, {Allen}, {Amici}, {Andersen}, {Antonelli}, {Antoniak}, {Attin{\`a}}, {Barbanera}, {Bachetti}, {Baggett}, {Bladt}, {Brez}, {Bonino}, {Boree}, {Borotto}, {Breeding}, {Brienza}, {Bygott}, {Caporale}, {Cardelli}, {Carpentiero}, {Castellano}, {Castronuovo}, {Cavalli}, {Cavazzuti}, {Ceccanti}, {Centrone}, {Citraro}, {D'Amico}, {D'Alba}, {Di Gesu}, {Del Monte}, {Dietz}, {Di Lalla}, {Persio}, {Dolan}, {Donnarumma}, {Evangelista}, {Ferrant}, {Ferrazzoli}, {Ferrie}, {Footdale}, {Forsyth}, {Foster}, {Garelick}, {Gunji}, {Gurnee}, {Head}, {Hibbard}, {Johnson}, {Kelly}, {Kilaru},
  {Lefevre}, {Roy}, {Loffredo}, {Lorenzi}, {Lucchesi}, {Maddox}, {Magazzu}, {Maldera}, {Manfreda}, {Mangraviti}, {Marengo}, {Marrocchesi}, {Massaro}, {Mauger}, {McCracken}, {McEachen}, {Mize}, {Mereu}, {Mitchell}, {Mitsuishi}, {Morbidini}, {Mosti}, {Nasimi}, {Negri}, {Negro}, {Nguyen}, {Nitschke}, {Nuti}, {Onizuka}, {Oppedisano}, {Orsini}, {Osborne}, {Pacheco}, {Paggi}, {Painter}, {Pavelitz}, {Pentz}, {Piazzolla}, {Perri}, {Pesce-Rollins}, {Peterson}, {Pilia}, {Profeti}, {Puccetti}, {Ranganathan}, {Ratheesh}, {Reedy}, {Root}, {Rubini}, {Ruswick}, {Sanchez}, {Sarra}, {Santoli}, {Scalise}, {Sciortino}, {Schroeder}, {Seek}, {Sosdian}, {Spandre}, {Speegle}, {Tamagawa}, {Tardiola}, {Tobia}, {Thomas}, {Valerie}, {Vimercati}, {Walden}, {Weddendorf}, {Wedmore}, {Welch}, {Zanetti}, \& {Zanetti}}]{Weisskopf2022}
{Weisskopf}, M.~C., {Soffitta}, P., {Baldini}, L., {et~al.} 2022, \bibinfo{title}{{The Imaging X-Ray Polarimetry Explorer (IXPE): Pre-Launch},} JATIS, 8, 026002, \dodoi{10.1117/1.JATIS.8.2.026002}

% type= article
\bibitem[{A.~A. {Zdziarski} {et~al.}(2013){Zdziarski}, {Mikolajewska}, \& {Belczynski}}]{Zdziarski2013}
{Zdziarski}, A.~A., {Mikolajewska}, J., \& {Belczynski}, K. 2013, \bibinfo{title}{{Cyg X-3: a low-mass black hole or a neutron star.},} \mnras, 429, L104, \dodoi{10.1093/mnrasl/sls035}

% type= article
\bibitem[{A.~A. {Zdziarski} {et~al.}(2016){Zdziarski}, {Segreto}, \& {Pooley}}]{Zdziarski2016}
{Zdziarski}, A.~A., {Segreto}, A., \& {Pooley}, G.~G. 2016, \bibinfo{title}{{The radio/X-ray correlation in Cyg X-3 and the nature of its hard spectral state},} \mnras, 456, 775, \dodoi{10.1093/mnras/stv2647}

% type= article
\bibitem[{A.~A. Zdziarski {et~al.}(2012)Zdziarski, Sikora, Dubus, Yuan, Cerutti, \& Ogorzałek}]{Zdziarski2012a}
Zdziarski, A.~A., Sikora, M., Dubus, G., {et~al.} 2012, \bibinfo{title}{The gamma‐ray emitting region of the jet in Cyg X‐3,} \mnras, 421, 2956, \dodoi{10.1111/j.1365-2966.2012.20519.x}

\end{thebibliography}
\bibliographystyle{aasjournalv7}



\end{document}